\documentclass[%
 reprint,
amsmath,amssymb,
aps,
]{revtex4-1}
\usepackage{graphicx}
\usepackage{dcolumn}
\usepackage{bm}
\usepackage[colorlinks, linkcolor=blue, anchorcolor=blue, citecolor=blue]{hyperref}
\usepackage{url}
\usepackage{amsmath,amssymb}
\usepackage{newtxmath}
\usepackage[mathlines]{lineno}
\usepackage{subfigure}
\usepackage{eurosym}
\usepackage{amsmath,amssymb,mathrsfs}
\usepackage{makecell}

\begin{document}

\preprint{APS/123-QED} 

\title{Engineering photonic band gaps with a waveguide-QED structure containing an atom-polymer array}
\author{M. S. Wang}
\affiliation{School of Physical Science and Technology, Southwest Jiaotong University, Chengdu 610031, China}
\author{W. Z. Jia}
\email{wenzjia@swjtu.edu.cn}
\affiliation{School of Physical Science and Technology, Southwest Jiaotong University, Chengdu 610031, China}

\date{\today}

\begin{abstract}
We investigate the generation and engineering of photonic band gaps in waveguide quantum 
electrodynamics systems containing periodically arranged atom-polymers. We first consider the configuration 
of a dimer array coupled to a waveguide. The results 
show that if the intra- and inter-cell phase delays are properly designed, the center and the 
width of the band gaps, as well as the dispersion relation of the passbands can be modified by adjusting the 
intra-cell coupling strength. These manipulations provide  ways to control the propagating modes in  the waveguide, leading to some interesting effects such as slowing or even stopping a single-photon pulse. Finally, we take the case of the tetramer chain as an example to show 
that, in the case of a larger number of atoms in each unit cell, tunable multi-gap structures and more 
sophisticated band-gap engineering can be realized. Our proposal provides efficient ways for photonic 
band-gap engineering in micro- and nano-quantum systems, which may facilitate the manipulation of photon transport in future quantum networks.
\end{abstract}

\maketitle


\section{\label{introduction}Introduction}

Waveguide quantum electrodynamics (wQED) systems are realized by strongly coupling quantum 
emitters (e.g., atoms, artificial atoms, cavities, and so on) to a one-dimensional (1D) waveguide and 
are ideal platform for the manipulation of strong emitter-photon interactions \cite{Roy-RMP2017, Gu-PhysReports2017,Sheremet-RMP2023}.
The high coupling efficiency between emitters and photonic modes makes wQED setups optimal for single-photon level quantum devices \cite{Shen-PRL2005,Chang-PRL2006,Shen-PRL2007,Zhou-PRL2008,Astafiev-Science2010,Shi-PRA2015}.
In a 1D wQED setup containing multiple emitters, the coherent and dissipative interactions between 
them can be mediated by their common electromagnetic environment, giving rise to a large number of 
intriguing phenomena, including superradiant and subradiant phenomena \cite{PR-Dicke1954,Vetter-PScr2016,Loo-Science2013,Zhang-PRL2019,Ke-PRL2019,Wang-PRL2020}, 
long-range quantum entanglement between atoms \cite{Zheng-PRL2013,Ballestero-PRA2014,Facchi-PRA-2016,Mirza-PRA2016},
cavity QED with atomic mirrors \cite{Chang-NJP2012,Mirhosseini-Natrue2019,Nie-PRL2023}, 
decoherence-free interactions in giant artificial atoms \cite{Kockum-PRL2018}, 
topologically protected scattering spectra \cite{Nie-PRApplied2021}, 
multiple Fano interferences \cite{Tsoi-PRA2008,Liao-PRA2015,Cheng-PRA2017,Ruostekoski-PRA2017,Mukhopadhyay-PRA2019}, 
control-field-free atomic coherence effects \cite{Mukhopadhyay-PRA2020,Ask-arXive2020,Jia-EPJP2022,Feng-PRA2021},
and so on.

In recent years, considerable attention has been devoted to photonic band-gap materials in micro- and 
nano-quantum structures. One potential avenue for achieving this objective is through the use of a 
conventional photonic crystal waveguide (PCW) \cite{Bendickson-PRE1996,Hood-PNAS2016},
in which the periodicity of the dielectric structure is employed to generate a set of Bloch bands. 
Moreover, in the microwave domain, PCW has a series of circuit versions, including, for example, 
coplanar waveguide with periodically modulated impedance \cite{Liu-NatPhys2017,Sundaresan-PRX2019}, resonator array waveguide \cite{Kim-PRX2021,Ferreira-PRX2021,Scigliuzzo-PRX2021,Zhang-Science2023},
and circuit-crystal metamaterials with Josephson junction arrays \cite{Rakhmanov-PRB2008,Hutter-PRB2011,Zueco-2PRB012}. 
Another method to generate photonic band gaps is based on wQED structures with periodically 
arranged emitters (either an atom array or a cavity array) side coupled to a 1D linear waveguide \cite{Xu-PRE2000,Shen-OL2005,Yanik-PRL2004,Shen-PRB2007,Yi-OE2010,Witthaut-NJP2010,Fang-PRA2015,Mirza-PRA2017,Mohammad-Natcom2018,He-OE2021,Greenberg-PRA2021,Tang-PRL2022,Peng-PRA2023,Berndsen-PRA2023,Berndsen-JOSAB2024}. 
Achieving highly tunable band-gap modulation in periodic structures that have already been prepared  
is of significant importance. This type of band-gap engineering can facilitate the manipulation of 
photon transport and the control of light-matter interactions in a variety of ways. In the case of band-gap structures produced by wQED containing multiple atoms, existing schemes for band-gap modulation are mainly based on the adjustment of the atomic frequency \cite{Shen-PRB2007,He-OE2021,Greenberg-PRA2021}
and the application of an external control field \cite{Witthaut-NJP2010}.

In micro- and nano-quantum systems, such as superconducting quantum circuits, the tunability of the 
coupling strength between artificial atoms is a prominent feature \cite{Hime-Science2006,Niskanen-Science2007,Baust-PRB2015,Zhu-PRA2019}. 
It seems reasonable to posit that the tunable coupling between atoms can be harnessed to engineer 
the band-gap structure formed by wQED systems containing an atomic array. In 
this paper we present a proposal to achieve this goal, as shown schematically in Fig.~\ref{Sketch}. In 
our proposal, a periodic atomic array is coupled to a 1D waveguide, with each unit cell 
containing at least two atoms with tunable couplings between the nearest neighboring atoms, forming 
a so called polymer. The transfer matrix of the system is derived using the real-space formalism, and 
then the corresponding scattering coefficients are expressed analytically in terms of Chebyshev 
polynomials. Using these results, we further analyze the band-gap structures in the scattering spectra and demonstrate the potential for band-gap engineering through the use of interatomic couplings.
Specifically, for the case where each cell is an atom-dimer (i.e., a cell with two coupled atoms), band-gap structures with highly tunable  center frequencies and widths can be realized by properly adjusting the intra-cell coupling strength and designing the  interatomic phase delays. Moreover, these kinds of modulations can also be used to engineer the propagating modes in the passband. In particular, if a band gap is closed 
at a specific frequency, the Bloch modes in the vicinity of this 
frequency will become slow modes with a linear dispersion relation. 
Furthermore, if the width of a passband is dynamically compressed by modulating the coupling strength, the formation of a flat band can enable the setup to slow down or even stop a single-photon pulse at will. 
Finally, the results for a dimer array can be generalized to the case of a polymer chain with a greater 
number of atoms in each unit cell. This enables the realization of a tunable multi-band-gap structure 
and the implementation of more sophisticated band-gap engineering.

The remainder of this paper is organized as follows. In Sec.~\ref{Model}, we present a theoretical 
model that includes the system Hamiltonian and corresponding single-photon transport equations. 
Furthermore, using the transfer matrix approach, we derive the expressions of the single-photon scattering coefficients. In Sec.~\ref{FormationBG}, we present some 
general discussions on the band-gap structure produced by a wQED system 
containing a polymerized atomic chain. In Sec.~\ref{DimerChain}, we analyze tunable band-gap structures 
and propagation-mode control in a wQED system 
coupled by an atom-dimer chain.
In Sec.~\ref{TetramerChain}, we further examine band-gap engineering in a wQED system 
containing an atom-tetramer chain.
Finally, further discussions and conclusions are given in Sec.~\ref{Conclusions}.
\section{\label{Model}Theoretical description of single-photon scattering}
\subsection{\label{HamiltonianAndEOM}Hamiltonian and single-photon transport equations}
\begin{figure}[t]
\centering
\includegraphics[width=0.5\textwidth]{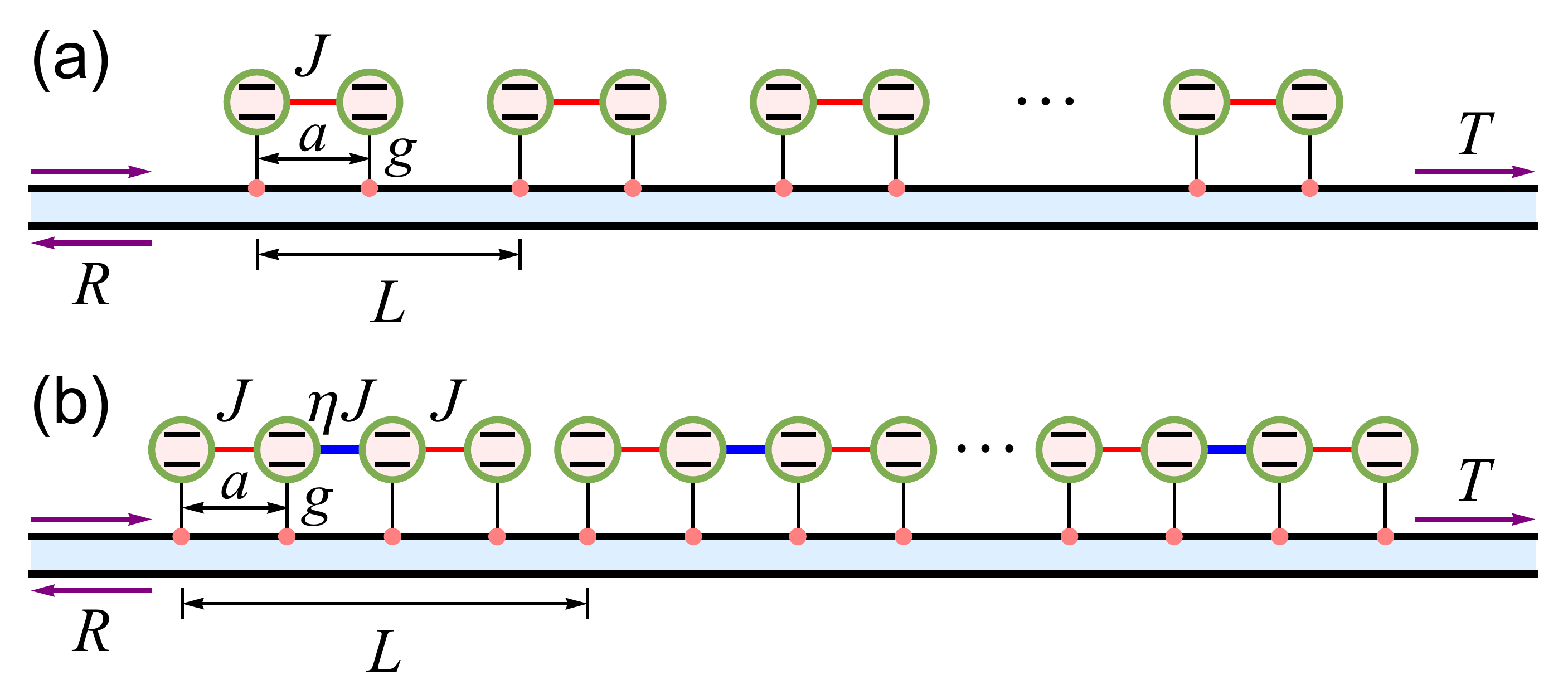}
\caption{Sketches of (a) an atom-dimer chain ($N=2$) coupled to a 1D waveguide, (b) an atom-tetramer chain ($N=4$) coupled to a 
1D waveguide.}
\label{Sketch}
\end{figure}

In this study, we concentrate on the wQED structures comprising $M$ periodically spaced polymers, each containing $N$ 
two-level atoms with identical transition frequency 
$\omega_{\mathrm{a}}$, coupled to photonic modes in a 1D waveguide with linear dispersion, as schematically shown in Fig.~\ref{Sketch}. 
The atoms are located at position $x_{m,s}$ ($m=1,2,\cdots,M$ labels the cells and $s=1,2,\cdots,N$ denotes the atoms
in each cell). The lattice constant of the polymer array is $L=x_{m,s}-x_{m-1,s}$. The distance between neighboring atoms in a cell
is $a=x_{m,s}-x_{m,s-1}$.
The strength between the $s$th atom
and the $(s+1)$th atom in a cell is $J_s$. The coupling strength between each atom and the waveguide is $g$. Under the rotating-wave approximation, 
the Hamiltonian of the system in the real space can be written as $(\hbar=1)$:
\begin{equation}
\begin{split}
\hat{H}=&\sum_{\rho}\int\mathrm{d}x{\hat{a}_{\rho}^{\dagger}(x)\left(-i\tau_{\rho}v_{\mathrm{g}}\frac{\partial}{\partial x} \right)\hat{a}_{\rho}\left( x \right)}
\\
&+\sum_{m=1}^{M}\sum_{s=1}^{N}\omega_{\mathrm{a}}\sigma_{m,s}^{+}\sigma_{m,s}^{-}+\sum_{m=1}^{M}\sum_{s=1}^{N-1}\left(J_{s}\sigma_{m,s}^{+}\sigma_{m,s+1}^{-}+\mathrm{H.c.}\right)
\\
&+\sum_{\rho}\sum_{m=1}^{M}\sum_{s=1}^{N}g\int \mathrm{d}x\delta(x-x_{m,s})\left[\hat{a}_{\rho}^{\dagger}\left(x\right)\sigma_{m,s}^{-}+\mathrm{H.c.}\right],
\end{split}
\label{Hamiltonian}
\end{equation}
where $\rho=\mathrm R, \mathrm L$
 and $\tau_{\mathrm R,\mathrm L} = \pm1$. $\hat{a}_{\mathrm{R}}^{\dagger}(x)$ [$\hat{a}_{\mathrm{R}}(x)$] and
$\hat{a}_{\mathrm{L}}^{\dagger}(x)$ [$\hat{a}_{\mathrm{L}}(x)$] are the bosonic creation (annihilation) operators of the right- and left-going wave at position $x$. $v_{\mathrm{g}}$ is the group velocity of photons in the waveguide.
$\sigma_{m,s}^{+}$ ($\sigma_{m,s}^{-}$) is the raising (lowering) operator of the $s$th atom in the $m$th cell. 

In the single-excitation manifold, the scattering eigenstate can be written as
\begin{equation}
|\Psi\rangle=\sum_{\rho}\int \mathrm{d}x\Phi_{\rho}\left(x\right)\hat{a}_{\rho}^{\dagger}\left(x\right)\left|\emptyset\right\rangle
+\sum_{m=1}^{M}\sum_{s=1}^{N}f_{m,s}\sigma_{m,s}^{+}\left|\emptyset\right\rangle,
\label{eigen state}
\end{equation}
where $|\emptyset\rangle$ is the vacuum state with no photons propagating in the waveguide 
and the atoms occupying their ground states. $\Phi_{\rho}(x)$ is the single-photon 
wave function in the $\rho$ mode. $f_{m,s}$ is the excitation amplitude of the $s$th atom 
in the $m$th cell. 

We assume that a single photon with energy $\omega=v_{\mathrm{g}} k$ is initially incident from the left, where $k$ is the wave vector of the photon. The corresponding right(left)-going wave function $\Phi_{\mathrm{R}}(x)$ [$\Phi_{\mathrm{L}}(x)$] takes the following \textit{Ansatz}:
\begin{subequations}
\begin{eqnarray}
\Phi_{\mathrm{R}}\left(x\right)&=&e^{ikx}\Big\{\theta\left(x_{1,1}-x\right)
\nonumber
\\
&&+\sum_{m=1}^{M}\sum_{s=1}^{N}t_{m,s}\left[\theta\left(x-x_{m,s}\right)-\theta\left(x-x_{m,s+1}\right)\right]\Big\},
\nonumber
\\
\label{PhWR}
\end{eqnarray}
\begin{equation}
\Phi_{\mathrm{L}}\left(x\right)=e^{-ikx}\sum_{m=1}^{M}\sum_{s=1}^{N}r_{m,s}\left[\theta\left(x_{m,s}-x\right)-\theta\left(x_{m,s-1}-x\right)\right].
\label{PhWL}
\end{equation}
\end{subequations}
Here, we define $x_{1,0}=-\infty$ and $x_{M,N+1}=+\infty$. $t_{m,s}$ ($r_{m,s}$) is the transmission (reflection) amplitude of the $s$th atom in the $m$th cell. And the transmission (reflection) amplitude of the last (first) atom $t_{M,N}\equiv t$ ($r_{1,1}\equiv r$) is defined as the transmission (reflection) amplitude of the atom array. $\theta(x)$ denotes the Heaviside step function. 

From the eigenequation $\hat{H}\left|\Psi\right\rangle=\omega\left|\Psi\right\rangle$, we can obtain the following set of equations:
\begin{subequations}
\begin{equation}
t_{m,s}-t_{m,s-1}+i\sqrt{\frac{\Gamma}{2}}\tilde{f}_{m,s}e^{-i\phi_{m,s}}=0,
\label{EoM1}
\end{equation}
\begin{equation}
r_{m,s+1}-r_{m,s}-i\sqrt{\frac{\Gamma}{2}}\tilde{f}_{m,s}e^{i\phi_{m,s}}=0,
\label{EoM2}
\end{equation}
\begin{eqnarray}
&&J_{s-1}\tilde{f}_{m,s-1}-\Delta\tilde{f}_{m,s}+J_{s}\tilde{f}_{m,s+1}
\nonumber
\\
&&+\frac{1}{2}\sqrt{\frac{\Gamma}{2}}\left[\left(t_{m,s}+t_{m,s-1}\right)e^{i\phi_{m,s}}
+\left(r_{m,s+1}+r_{m,s}\right)e^{-i\phi_{m,s}}\right]=0.
\nonumber
\\
\label{EoM3}
\end{eqnarray}
\end{subequations}
Note that in the above equations the inter-cell connection relations $t_{m,0}=t_{m-1,N}, r_{m,N+1}=r_{m+1,1}$ and the boundary
conditions $t_{1,0}=1, r_{1,1}=r, t_{M,N}=t, r_{M,N+1}= 0$ are satisfied. $\tilde{f}_{m,s}=f_{m,s}/\sqrt{v_{\mathrm{g}}}$ is the atomic 
excitation amplitude scaled by $\sqrt{v_{\mathrm{g}}}$. $\Gamma=2g^{2}/v_{\mathrm{g}}$ is the decay rate of the atoms into the guided modes.  $\Delta=\omega-\omega_{\mathrm{a}}$ is the detuning between the photon and the atom. The phase factor corresponding to position $x_{m,s}$ is 
defined as $\phi_{m,s}=kx_{m,s}$.
\subsection{\label{TransferAndSC}Transfer matrix and scattering coefficients}
From Eqs.~\eqref{EoM1} and \eqref{EoM2}, after iteration, we can obtain
\begin{subequations}
\begin{equation}
t_{m,s}=t_{m-1,N}-i\sqrt{\frac{\Gamma}{2}}\sum_{j=1}^{s}e^{-i\phi_{m,j}}\tilde{f}_{m,j},
\label{tms}
\end{equation}	
\begin{equation}
r_{m,s}=r_{m+1,1}-i\sqrt{\frac{\Gamma}{2}}\sum_{j=s}^{N}e^{i\phi_{m,j}}\tilde{f}_{m,j}.
\label{rms}
\end{equation}
\end{subequations}
From now on, we rewrite the phase factor of the $s$th atom in the $m$th cell as $\phi_{m,s}=\beta_{m}+\phi_{s}$. $\beta_m=(m-1)\beta$ 
is the reference phase of the $m$th cell (without loss of generality, we have chosen $\beta_1=0$), where 
$\beta=\phi_{m,s}-\phi_{m-1,s}=kL$ is the phase delay between neighboring cells. $\phi_{s}=[s-(N+1)/2]\alpha$ is the relative phase 
of the $s$th atom in each cell with respect to the reference phase, where 
$\alpha=\phi_{m,s}-\phi_{m,s-1}=ka$ is the phase delay between neighboring atoms in a unit cell. 
Note that the phase delays 
\begin{equation}
\alpha=ka=\left(1+\frac{\Delta}{\omega_{\mathrm{a}}}\right)\alpha_0,~~\beta=kL=\left(1+\frac{\Delta}{\omega_{\mathrm{a}}}\right)\beta_0
\label{alphabeta}
\end{equation}
is detuning dependent, where $\alpha_0=\omega_{\mathrm{a}}a/{v_{\mathrm{g}}}$ and $\beta_0=\omega_{\mathrm{a}}L/{v_{\mathrm{g}}}$
are detuning-independent phase delays defined in terms of the frequency of the resonant photons.

Using the above definitions and substituting Eqs.~\eqref{tms} and \eqref{rms} into Eq.~\eqref{EoM3}, we have
\begin{eqnarray}
\left(\Delta\mathbf{I}-\mathbf{H_c}\right)\tilde{\mathbf{f}}_m=e^{i\beta_m}{\mathbf{V}}t_{m-1,N}+e^{-i\beta_m}{\mathbf{V}}^*r_{m+1,1},
\label{LEf}
\end{eqnarray}
where  $\mathbf{I}$ is the $N$-dimensional identity matrix. $\mathbf{H_{c}}$ is the effective non-Hermitian Hamiltonian of each cell, with elements
\begin{equation}
\left(\mathcal{H}_{\mathrm{c}}\right)_{ss'}=J_{s}\delta_{s,s'-1}+J_{s'}\delta_{s-1,s'}-i\frac{\Gamma}{2}e^{i\left|\phi_{s}-\phi_{s'}\right|}.
\label{EffHc}
\end{equation}
$\mathbf{V}$ and $\tilde{\mathbf{f}}_m$ take the form 
\begin{subequations}
\begin{equation}
\mathbf{V}=\sqrt{\frac{\Gamma}{2}}\left(e^{i\phi_1},e^{i\phi_2},\cdots,e^{i\phi_{N}}\right)^{\top},
\label{VNmatrix}
\end{equation}
\begin{equation}
\tilde{\mathbf{f}}_m=\begin{pmatrix}\tilde{f}_{m,1},&\tilde{f}_{m,2},&\cdots,&\tilde{f}_{m,N}\end{pmatrix}^{\top}.
\label{fNmatrix}
\end{equation}
\end{subequations}

According to Eqs~\eqref{tms} and \eqref{rms}, we can read the relations satisfied by the transmission (reflection) amplitudes of neighboring cells: 
\begin{subequations}
\begin{equation}
t_{m,N}=t_{m-1,N}-ie^{-i\beta_{m}}{\mathbf{V}}^{\dagger}\tilde{\mathbf{f}}_{m},
\label{tmN}
\end{equation}
\begin{equation}
r_{m,1}=r_{m+1,1}-ie^{i\beta_m}{\mathbf{V}}^{\top}\tilde{\mathbf{f}}_m.
\label{rm1}
\end{equation}
\end{subequations}
On the other hand, Eq.~\eqref{LEf} allows us to express the vector $\tilde{\mathbf{f}}_m$, which is composed of the 
excitation amplitudes of the atoms in the $m$th cell, in terms of the corresponding input amplitudes 
$t_{m-1,N}, r_{m+1,1}$. Substituting this result into Eqs.~\eqref{tmN} and \eqref{rm1}, one can 
obtain the connection relation between the 
input and the output amplitudes of the $m$th unit cell
\begin{equation}
\binom{t_{m,N}}{r_{m,1}}=\mathscr{B}^{-1}_{m}\mathcal{S}\mathscr{B}_{m}\binom{t_{m-1,N}}{r_{m+1,1}},
\label{SConnect}
\end{equation}
where $\mathscr{B}_{m}=\mathrm{diag}(e^{i\beta_m},e^{-i\beta_m})$ and the scattering matrix $\mathcal{S}$ takes the form
\begin{equation}
\renewcommand{\arraystretch}{1.5}
\mathcal{S}=
\begin{pmatrix}
{1-i\mathbf{V}^{\dagger}(\Delta\mathbf{I}-\mathbf{H}_{\mathbf{c}})^{-1}\mathbf{V}}
&{-i\mathbf{V}^{\dagger}(\Delta\mathbf{I}-\mathbf{H}_{\mathbf{c}})^{-1}\mathbf{V}^{*}}
\\
{-i\mathbf{V}^{\top}(\Delta\mathbf{I}-\mathbf{H}_{\mathbf{c}})^{-1}\mathbf{V}}
&{1-i\mathbf{V}^{\top}(\Delta\mathbf{I}-\mathbf{H}_{\mathbf{c}})^{-1}\mathbf{V}^{*}}
\end{pmatrix}.
\label{Smatrix}
\end{equation}
Alternatively, the connection relation can be expressed as 
\begin{equation}
\binom{t_{m-1,N}} {r_{m,1}}=\mathscr{B}^{-1}_{m}\mathcal{T}\mathscr{B}_{m}\binom{t_{m,N}} {r_{m+1,1}}.
\label{TConnect1}
\end{equation}
Here 
\begin{equation}
\renewcommand{\arraystretch}{2}
\mathcal{T}=
\begin{pmatrix}
\frac{1}{1-i\mathbf{V}^{\dagger}(\Delta\mathbf{I}-\mathbf{H}_{\mathrm{c}})^{-1}\mathbf{V}}
&\frac{i\mathbf{V}^{\dagger}(\Delta\mathbf{I}-\mathbf{H}_{\mathrm{c}})^{-1}\mathbf{V}^{*}}{1-i\mathbf{V}^{\dagger}(\Delta\mathbf{I}-\mathbf{H}_{\mathrm{c}})^{-1}\mathbf{V}}
\\
\frac{-i\mathbf{V}^{\top}(\Delta\mathbf{I}-\mathbf{H}_{\mathrm{c}})^{-1}\mathbf{V}}{1-i\mathbf{V}^{\dagger}(\Delta\mathbf{I}-\mathbf{H}_{\mathrm{c}})^{-1}\mathbf{V}}
&\frac{\mathrm{Det}[\mathcal{S}]}{1-i\mathbf{V}^{\dagger}(\Delta\mathbf{I}-\mathbf{H}_{\mathrm{c}})^{-1}\mathbf{V}}
\end{pmatrix}
\label{Tmatrix1}
\end{equation}
represents the transfer matrix, which establishes a relationship between the light field on either side of the $m$th unit cell. As indicated by Eq.~\eqref{TConnect1}, the connection 
relation for a single-polymer setup (i.e., the case of $M=1$) with the reference phase set to zero acquires the following form 
\begin{equation}
\left(\begin{array}{c}1\\ \tilde{r}\end{array}\right)
=\mathcal{T}\left(\begin{array}{c}\tilde{t}\\0\end{array}\right),
\label{TConnect2}
\end{equation}
where $\tilde{t}$ and $\tilde{r}$ are the corresponding transmission and reflection amplitudes. 
Furthermore, by inserting Eq.~\eqref{Tmatrix1} into Eq.~\eqref{TConnect2}, it can be observed that
\begin{subequations}
\begin{equation}
\tilde{t}=1-i\mathbf{V}^{\dagger}(\Delta\mathbf{I}-\mathbf{H}_{\mathrm{c}})^{-1}\mathbf{V},
\label{tilde_t}
\end{equation}
\begin{equation}
\tilde{r}=-i\mathbf{V}^{\top}(\Delta\mathbf{I}-\mathbf{H}_{\mathrm{c}})^{-1}\mathbf{V}.
\label{tilde_r}
\end{equation}
\end{subequations}
In addition, the atomic chain under consideration satisfies time reversal symmetry. Consequently, 
the transfer matrix can also be expressed in terms of $\tilde{t}$ and $\tilde{r}$ as follows \cite{Xu-PRE2000} 
\begin{equation}
\renewcommand{\arraystretch}{1.5}
\mathcal{T}=
\begin{pmatrix}
\frac{1}{\tilde{t}}&\frac{\tilde{r}^*}{\tilde{t}^*}
\\
\frac{\tilde{r}}{\tilde{t}}&\frac{1}{\tilde{t}^*}
\end{pmatrix}.
\label{Tmatrix2}
\end{equation}

Using Eq.~\eqref{TConnect1} iteratively $M$ times in succession and noticing the conditions $t_{0,N}=1$, $r_{1,1}=r$, $t_{M,N}=t$, $r_{M+1,1}=0$, and $\beta_1=0$, we finally obtain the following connection relation between the reflection and transmission amplitudes for the entire atomic chain:
\begin{equation}
\left(\begin{array}{c}1\\r\end{array}\right)=\tilde{\mathcal{T}}^M\left(\begin{array}{c}te^{iM\beta}\\0\end{array}\right),
\label{simultaneous equation1}
\end{equation}
with
\begin{equation}
\tilde{\mathcal{T}}=\mathcal{T}\mathcal{T}_\beta,~~\mathcal{T}_\beta=\begin{pmatrix}e^{-i\beta}&0
\\0&e^{i\beta}\end{pmatrix}.
\label{Tbeta}
\end{equation}
Thus the transmission and reflection amplitudes for the polymer chain read
\begin{equation}
t=\frac{e^{-iM\beta}}{[\tilde{\mathcal{T}}^M]_{11}},~~r=\frac{[\tilde{\mathcal{T}}^M]_{21}}{[\tilde{\mathcal{T}}^M]_{11}},
\label{TotalAmp}
\end{equation}	

Notice that the determinant of the matrix $\tilde{\mathcal{T}}$ is $1$. Based on Abeles's theorem \cite{Abeles-AnnPhys1950}, 
we can write the matrix $\tilde{\mathcal{T}}^{M}$ in terms of Chebyshev polynomials of the second kind:
\begin{equation}
\tilde{\mathcal{T}}^{M}=U_{M-1}(y)\tilde{\mathcal{T}}-U_{M-2}(y)\mathcal{I},
\label{decomposeTM}
\end{equation}
with
\begin{equation}
y=\frac{1}{2}\mathrm{Tr}\left(\tilde{\mathcal{T}}\right).
\label{y}
\end{equation}
$\mathcal{I}$ is the two-dimensional identity matrix. According to Eqs.~\eqref{TotalAmp} and \eqref{decomposeTM}, one can 
obtain the transmittance and the reflectance of the polymer chain (see Appendix \ref{DerivationEQ23} for details)
\begin{subequations}
\begin{equation}
T=\left|t\right|^2=\frac{1}{1+\zeta U_{M-1}^2(y)},
\label{Transmission}
\end{equation}	
\begin{equation}
R=\left|r\right|^2=\frac{\zeta U_{M-1}^2(y)}{1+\zeta U_{M-1}^2(y)},
\label{Reflection}
\end{equation}
\end{subequations}
where $\zeta\equiv|\tilde{\mathcal{T}}_{12}|^{2}=|\tilde{\mathcal{T}}_{21}|^{2}$.  It should be pointed out that the relation $T+R=1$ is satisfied as a consequence of the 
conservation of photon number. Therefore, in what follows, we will focus on the reflectance $R$ alone when analyzing the band-gap structures.

Moreover, we will focus on the regime where the phase-accumulated effects for detuned photons can be ignored, i.e., 
one can approximate the detuning-dependent phase delays defined in Eq.~\eqref{alphabeta} as $\alpha\simeq\alpha_0$ and $\beta\simeq\beta_0$. Or 
equivalently, we can say that the Markovian approximation can be performed, i.e., the time delays between the coupling points can be neglected. 
In Appendix~\ref{MarkovCond}, we present the Markovian conditions for our system, which are easily satisfied under typical parameters 
of wQED experiments \cite{Sheremet-RMP2023}.
In what follows, we will use this approximation in our analysis to keep the physics transparent, while in the full numerical calculations we still make the 
phase delay $\alpha$ and $\beta$ depend on the detuning in order to obtain accurate results.
\begin{figure}[t]
\centering
\includegraphics[width=0.5\textwidth]{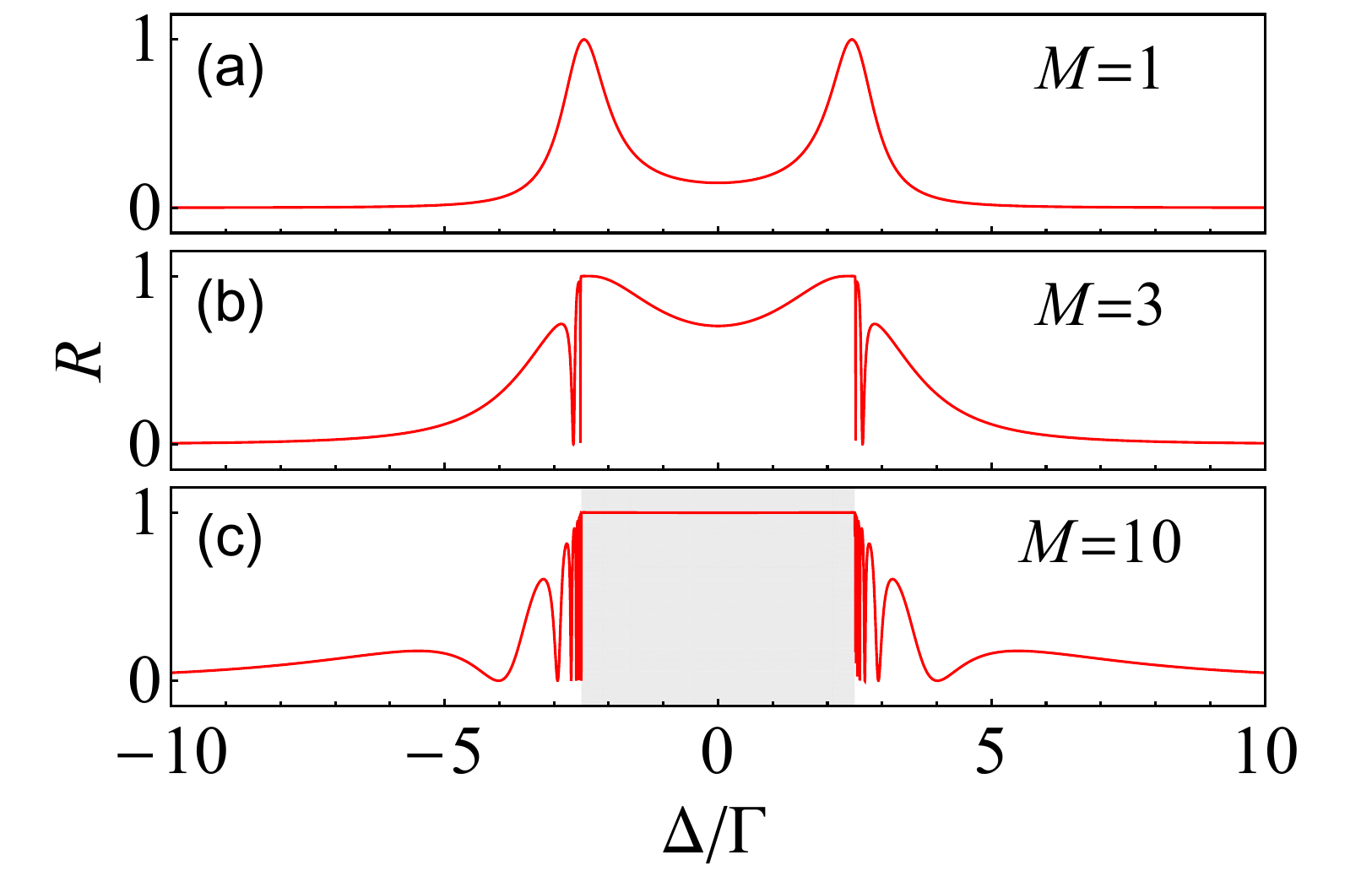}
\caption{Reflection spectra for a dimer chain with (a) $M=1$, (b) $M=3$ and (c) $M=10$. The band gap is represented by the shaded region.
Other parameters are set as $\alpha_0=\pi/2$, $\beta_0=\pi$, $J=2\Gamma$, and $\omega_{\mathrm{a}}=10^{4}\Gamma$.}
\label{BandgapForm}
\end{figure}
\begin{figure}[t]
\centering
\includegraphics[width=0.5\textwidth]{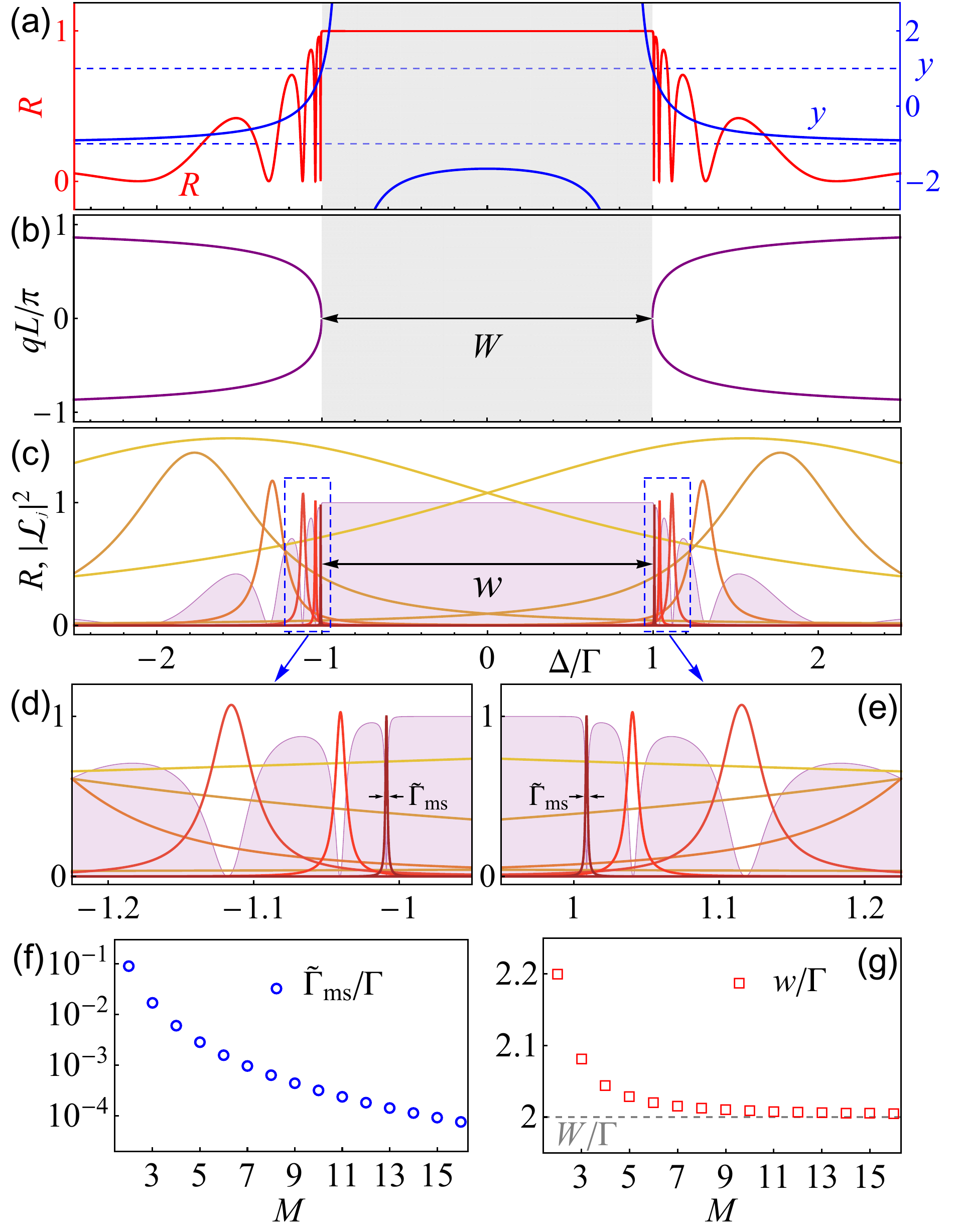}
\caption{(a) The reflection coefficient $R$ and $y$ as functions of $\Delta$ for an atom-dimer chain with $6$ cells. 
The blue dashed lines are used to mark $y=\pm 1$. 
The parameters are set as $\alpha_0=\pi/2$, $\beta_0=\pi$, $J=\Gamma/2$, and $\omega_{\mathrm{a}}=10^4\Gamma$.
(b) The dispersion curve for a dimer chain with the same parameters used in panel (a). In panels (a) and (b), 
the band gap of width $W$ is represented by the shaded region (with $|y|>1$). 
(c) The curves show the decomposed components contributed by the collective excitations corresponding to the spectra 
shown in panel (a). $w$ represents the distance between the two narrowest peaks,  which are located at the 
innermost and correspond to the two most subradiant states. (d), (e) Details of the curves in the blue dashed boxes in panel (c).  
$\tilde{\Gamma}_{\mathrm{ms}}$ denotes the width of the narrowest peaks. 
(f) The blue circles illustrate the width $\tilde{\Gamma}_{\mathrm{ms}}$ as a function of $M$. (g) The red squares show the distance 
$w$ as a function of $M$. As the value of $M$ increases, the value of $w$ approaches $W$.
Other parameters used in panels (f) and (g) are the same as those used in panel (a).}
\label{CMBG}
\end{figure}
\section{\label{FormationBG}Formation of band-gap structure}
In this section, we present some general discussions on the band-gap structure produced by a polymerized atomic chain.  
The existence of band gaps is a common feature of periodic structures. For our
system, it is not surprising that band-gap structures can form as the number of cells $M$ increases.
To illustrate this point, we present in Figs.~\ref{BandgapForm}(a)-\ref{BandgapForm}(c) the reflection spectra for an atom-dimer array for varying 
values of $M$. As $M$ increases, the spectrum with two total reflection points for small $M$ evolves 
into a band-gap structure for relatively large $M$. Moreover, it can be observed from
Eq.~\eqref{Reflection} that the solutions to the equation for $\Delta$
\begin{equation}
y(\Delta)=\cos\left(\frac{l\pi}{M}\right)\equiv y_l
\label{ys}
\end{equation}
are the zero points of the function $R(\Delta)$, since $y_l$ ($l=1,2,\cdots,M-1$) is a root of the the Chebyshev polynomial of the second kind
$U_{M-1}(y)$ \cite{Greenberg-PRA2021,Peng-PRA2023}. Importantly, these zero reflection points 
outside the band gap indicate the frequencies of the propagating modes that are permitted  
by the passbands.

In order to gain insight into the formation of the band gap, we consider a periodic chain comprising an 
adequate number of cells and present the dispersion relation based on Bloch's theorem (see 
Appendix \ref{DerivationDR} for details):
\begin{equation}
\cos\left(qL\right)=y\left(\Delta\right).
\label{DispersionR}
\end{equation}
Here $q$ represents the Bloch wave number, which is used to label the propagating 
modes that are permitted by the wQED system with atomic chain. Clearly, Eq.~\eqref{DispersionR} tells us that for 
frequencies satisfying the inequality 
\begin{equation}
|y(\Delta)|>1,
\label{BGConditionY}
\end{equation}
no propagating mode exists, resulting in band-gap structures with $R=1$ in the 
reflection spectrum, as shown by the shaded region in Fig.~\ref{CMBG}(a). The corresponding dispersion relation, which can only
be defined in the pass bands, is shown in Fig.~\ref{CMBG}(b). 
 
On the other hand, the formation of a spectrum with band-gap structures can be understood in terms of interference effects between 
different scattering channels corresponding to the excitations of the collective modes. To this end, we decompose the reflection spectrum into the 
superpositions of several Lorentzian-type amplitudes contributed by the collective excitations (see Appendix \ref{CollectiveModes} for details):
\begin{equation}
r=\sum_{j=1}^{N'}\mathcal{L}_j,~~\mathcal{L}_j=\frac{d_j}{\Delta-\tilde{\Delta}_j+i\frac{\tilde{\Gamma}_j}2}.
\label{rDecompose2}
\end{equation}	
Here $N'=MN$ is the total number of atoms, $\tilde{\Delta}_j$ represents the detuning between the $j$th collective mode and the atoms,
$\tilde{\Gamma}_{j}$ denotes the effective decay of the $j$th collective mode, and $d_{j}$ determines the weight of each Lorentzian 
component.  
In Figs.~\ref{CMBG}(c)-\ref{CMBG}(e), we illustrate this decomposition by taking the case of a dimer array
(with $\alpha_0=\pi/2,~\beta_0=\pi$) as an example.
It can can be seen that the Lorentzian-type amplitudes $\mathcal{L}_{j}$ are symmetrically distributed around $\Delta=0$. The band-gap region with near total reflection is mainly attributed to the excitation of the most superradiant states [see Fig.~\ref{CMBG}(c)]. While the two narrowest excitation amplitudes 
corresponding to the most subradiant states are situated at the innermost [see Figs.~\ref{CMBG}(d) and \ref{CMBG}(e)], and the corresponding width, 
denoted as $\tilde{\Gamma}_{\mathrm{ms}}$, decreases sharply with increasing number of cells $M$ [see Fig.~\ref{CMBG}(f)]. 
The Fano-type destructive interferences between these two most subradiant excitations and the superradiant ones result in the formation of 
a pair of innermost reflection minima and the corresponding steep band-gap walls [see Figs.~\ref{CMBG}(c)-\ref{CMBG}(e)]. This phenomenon becomes more remarkable as the parameter $M$ increases. Consequently, as the value of $M$ increases to approximately
$M>5$, the distance $w$ between the centers of the two innermost amplitudes approaches the width of the band gap $W$ [which is 
obtained from the condition \eqref{BGConditionY}], as shown in Fig.~\ref{CMBG}(g). This result indicates that the condition \eqref{BGConditionY} can even be used to calculate the gap width of a chain containing a relatively small number of cells.
%
%
\section{\label{DimerChain}Tunable photonic band gap with a dimer chain}
Let us first consider the simplest case of $N=2$, namely a dimer array coupled to a waveguide, 
as shown in Fig.~\ref{Sketch}(a). 
As demonstrated in the preceding sections, once the expressions for the quantities $\zeta$ and $y$ 
[from which we can determine the scattering coefficients \eqref{Transmission} and \eqref{Reflection}, 
as well as the dispersion relation \eqref{DispersionR}]
have been derived, a more comprehensive examination of the scattering spectrum and the band-gap structure is possible. Specifically, in the Markovian regime with $\alpha\simeq\alpha_{0}$ and $\beta\simeq\beta_{0}$, the quantities $\zeta$ and $y$ for a dimer array take the form 
\begin{subequations}
\begin{equation}
\zeta=\frac{\Gamma^2\left(2\Delta\cos\alpha_0+2J+\Gamma\sin\alpha_{0}\right)^2}{4\left[\Delta^{2}-J(J+\Gamma\sin\alpha_{0})\right]^2},
\label{xidimer}
\end{equation}	
\begin{eqnarray}
y&=&\frac{\frac{1}{2}\Gamma^{2}\sin(\beta_{0}-\alpha_{0})\sin\alpha_{0}+\Gamma(J\cos\alpha_{0}+\Delta)\sin\beta_{0}}{\Delta^{2}-J(J+\Gamma\sin\alpha_{0})}
\nonumber
\\
&&+\cos\beta_0,
\label{ydimer}
\end{eqnarray}
\end{subequations}
%
which are strongly influenced by the intra-cell phase difference $\alpha_{0}$ and the
inter-cell phase differences $\beta_{0}$.
Here we focus on the cases that the intra- and inter-cell phase delays satisfy either the Bragg or the anti-Bragg condition. The following analysis will demonstrate that, with the exception of the case where both the intra- and inter-cell phase delays satisfy the Bragg condition (see Sec.~\ref{BvsB}), spectra with highly tunable band-gap structures can be generated in the remaining cases by varying the intra-cell coupling strength $J$ (see Secs.~\ref{ABvsB}-\ref{ABvsAB}).
\begin{figure}[t]
\centering
\includegraphics[width=0.5\textwidth]{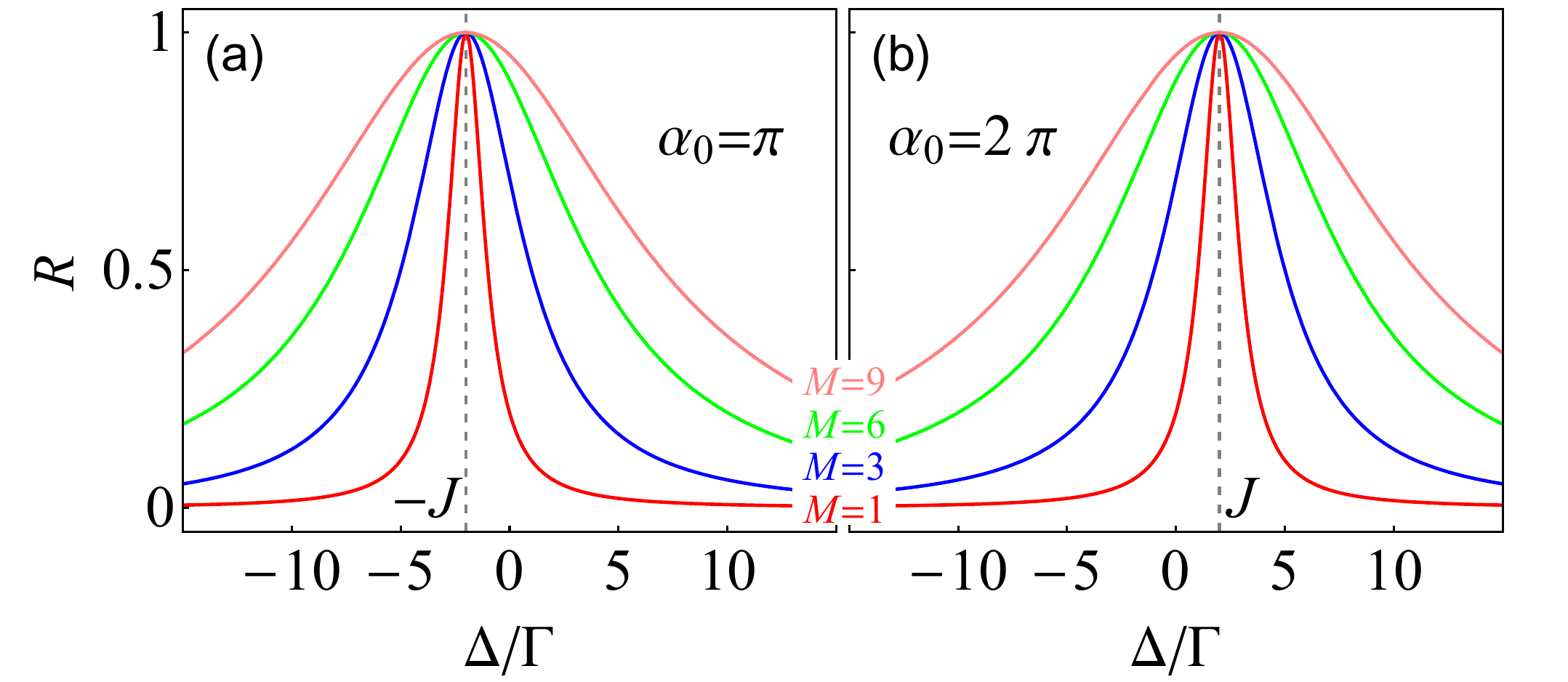}
\caption{Reflection spectra for a dimer chain with (a) $\alpha_{0}=\pi$ and (b) $\alpha_{0}=2\pi$. 
Other parameters are set as $\beta_{0}=4\pi$, $J=2\Gamma$, and $\omega_{\mathrm{a}}=10^{5}\Gamma$.}
\label{LorentzianSP}
\end{figure}
\subsection{\label{BvsB}$\alpha_{0}=n\pi$ and $\beta_{0}=m\pi$: spectrum with Lorentzian line shape} 
We begin with the assumption that both the intra- and inter-cell phase differences satisfy the Bragg condition, with $\alpha_{0}=n\pi$ ($n\in\mathbb{N}^{+}$) and $\beta_{0}=m\pi$ ($m\in\mathbb{N}^{+}$, $m>n$). Accordingly, Eqs.~\eqref{xidimer} and \eqref{ydimer}
can be simplified as
\begin{subequations}
\begin{equation}
\zeta=\frac{\Gamma^2}{\left[\Delta-(-1)^nJ\right]^2},
\label{xidimer1}
\end{equation}	
\begin{equation}
y=(-1)^m.
\label{ydimer1}
\end{equation}
\end{subequations}
Plugging these expressions into Eq.~\eqref{Reflection} and using the identity $U^2_{M-1}[(-1)^{m}]=M^2$,  the reflectance can be further written as
\begin{equation}
R=\frac{M^2\Gamma^2}{\left[\Delta-(-1)^nJ\right]^2+M^2\Gamma^2}.
\label{Rcase1}
\end{equation}
Obviously, similar to an uncoupled atom array (with $J=0$) under Bragg condition, the reflection spectrum shows the well-known phenomenon of superradiance, with a line width $2M\Gamma$ (scaled as the number of atoms $2M$). However, the center frequency is shifted to $\Delta=(-1)^{n}J$
instead of at $\Delta=0$. 
These properties are shown in Figs.~\ref{LorentzianSP}(a) and \ref{LorentzianSP}(b), where we take the case of $\alpha_{0}=\pi$ ($n=1$) and $\alpha_{0}=2\pi$ ($n=2$) as examples. 

\begin{figure}[t]
\centering
\includegraphics[width=0.5\textwidth]{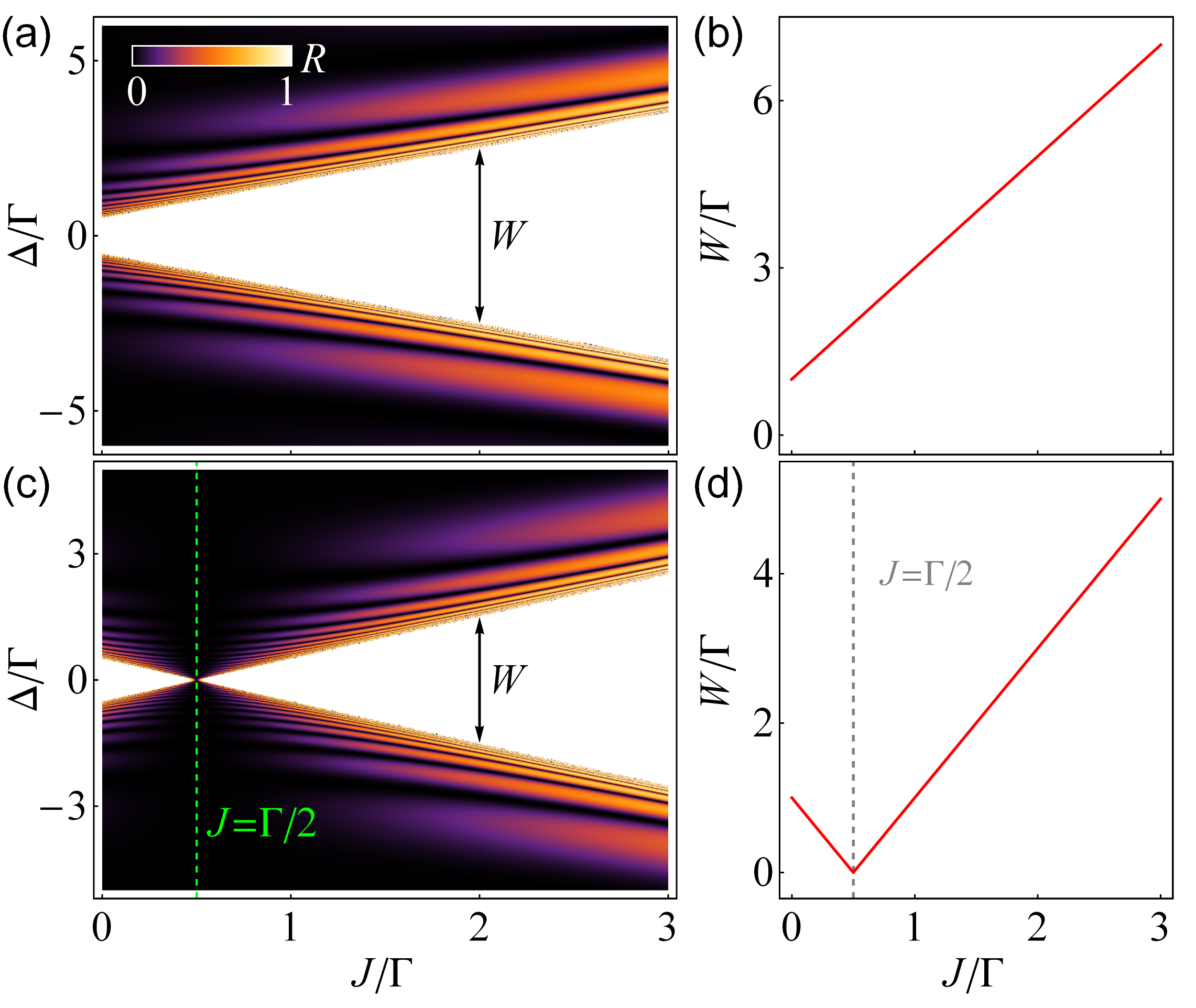}
\caption{(a) Reflectance for a dimer chain as a function of $\Delta$ and $J$. The parameters
are set as $M=15$, $\alpha_{0}=\pi/2$, $\beta_0=3\pi$, and $\omega_{\mathrm{a}}=10^{4}\Gamma$, respectively.
(b) The width $W$ of the band gap as a function of $J$ corresponding to the case shown in panel (a). 
(c) Reflectance for a dimer chain as a function of $\Delta$ and $J$ with intra-cell phase delay $\alpha_{0}=3\pi/2$. Other parameters are the same as those used in panel (a).
(d) The width $W$ as a function of $J$ corresponding to the case shown in panel (c). 
The dashed lines in panels (c) and (d) are employed to indicate the value of $J$ corresponding to $W=0$.}
\label{DimerBGT1}
\end{figure}
\begin{figure}[t]
\centering
\includegraphics[width=0.5\textwidth]{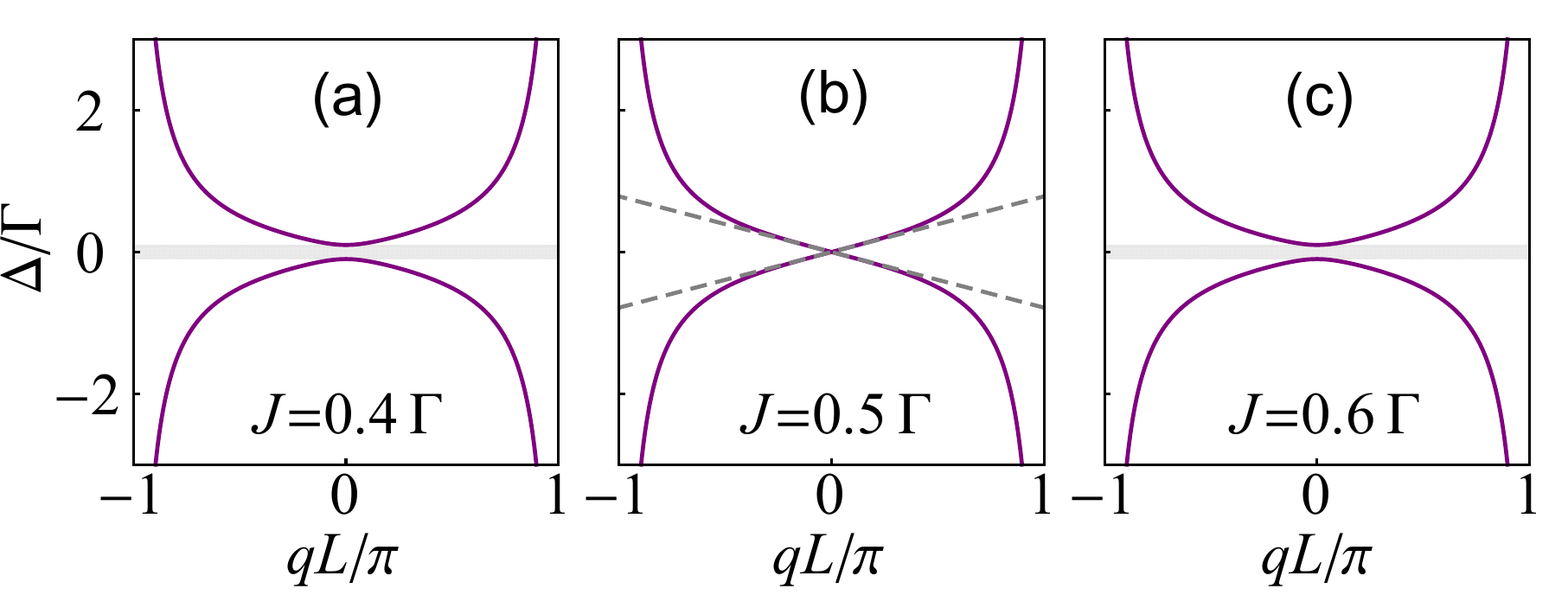}
\caption{Dispersion relations for a dimer chain for various values of $J$ in the vicinity of $J=\Gamma/2$. 
The parameters are the same as those used in Fig.~\ref{DimerBGT1}(c).}
\label{DimerDisRela1}
\end{figure}
\subsection{\label{ABvsB}$\alpha_{0}=(n-1/2)\pi$ and $\beta_{0}=m\pi$: band gap with tunable width}
We proceed to assume that $\alpha_{0}=(n-1/2)\pi$ ($n\in\mathbb{N}^{+}$) and $\beta_{0}=m\pi$ ($m\in\mathbb{N}^{+}$, $m\geq n$), i.e., 
the intra(inter)-cell phase difference satisfies the anti-Bragg (Bragg) condition.
Under these conditions, Eqs.~\eqref{xidimer} 
and \eqref{ydimer} can be simplified as
\begin{subequations}
\begin{equation}
\zeta=\frac{\Gamma^{2}\left[2J-(-1)^n\Gamma\right]^{2}}{4\left\{\Delta^{2}-J\left[J-(-1)^n\Gamma\right]\right\}^{2}},
\label{xidimer2}
\end{equation}	
\begin{equation}
y=(-1)^m\left(1-\frac{\frac{1}{2}\Gamma^2}{\Delta^2-J\left[J-(-1)^n\Gamma\right]}\right).
\label{ydimer2}
\end{equation}
\end{subequations}
After substituting Eq.~\eqref{ydimer2} into the band-gap condition \eqref{BGConditionY}, we find that except for the case of $n\in\mathbb{E}^{+},J=\Gamma/2$ (where the gap is closed), the region of the band gap is
\begin{equation}
|\Delta|<|J\pm\Gamma/2|. 
\label{BGrange1}
\end{equation}
Namely, we can obtain a single band gap centered about $\Delta=0$, with width
\begin{equation}
W=\left|2J\pm\Gamma\right|.
\label{Wcase1}
\end{equation}
Here ``$+$'' (``$-$'') corresponds to $n\in\mathbb{O}^{+}$ ($n\in\mathbb{E}^{+}$). It should be emphasized that the value of $m$ do not influence the solution set of the inequality \eqref{BGConditionY}, thus the range and width of the band gap are only dependent on the parity of $n$. A similar conclusion can be drawn with regard to the situations discussed in the following subsections (Sec.~\ref{BvsAB} and \ref{ABvsAB}). Importantly, the results obtained in Eqs.~\eqref{BGrange1} and \eqref{Wcase1} indicate that the width of the band gap can be manipulated by changing the coupling strength $J$ between the atoms 
within a cell, which is a significant advantage over a chain of non-interacting atoms.

Without loss of generality, we will consider the cases of $\alpha_{0}=\pi/2$ ($n=1$) and $\alpha_{0}=3\pi/2$ ($n=2$) 
as illustrative examples to demonstrate the modulation effect of $J$ on the width of the band gap.
To this end, we plot the reflection spectra and corresponding gap width under 
these phase delays for varying $J$ in Fig.~\ref{DimerBGT1}.  Specifically, when $\alpha_{0}=\pi/2$ and $\beta_{0}=3\pi$, the bandwidth increases 
linearly with $J$. Particularly, if $J\gg\Gamma$, we can obtain a wide band gap with width $W\simeq2J$ 
[see Figs.~\ref{DimerBGT1}(a) and \ref{DimerBGT1}(b)]. On the other hand, when $\alpha_{0}=3\pi/2$ 
and $\beta_{0}=3\pi$, the bandwidth $W$ first decreases linearly, reaching a minimum value of zero 
at $J=\Gamma/2$, and then increases linearly with $J$ to a wide band gap [see Figs.~\ref{DimerBGT1}(c) and \ref{DimerBGT1}(d)].
Additionally, the dispersion relations for this case around $J=\Gamma/2$ are plotted in Figs.~\ref{DimerDisRela1}(a)-\ref{DimerDisRela1}(c).
It is shown that at $J=\Gamma/2$ the gap is closed, so the atom-coupled waveguide, like a bare waveguide, 
allows photons of all frequencies [see Fig.~\ref{DimerDisRela1}(b)].
But its dispersion relation has been greatly modified compared to a bare waveguide. 
In particular, around the resonance point $\Delta=0$, one can 
obtain an effective linear dispersion $|\Delta|\simeq\tilde{v}_{\mathrm{g}}|q|$, with $\tilde{v}_{\mathrm{g}}=\Gamma L/4$ [see the 
dashed lines in Fig.~\ref{DimerDisRela1}(b)]. 
Moreover, the ratio between the group velocities of the perturbed and the bare waveguides can be estimated as 
$\tilde{v}_{\mathrm{g}}/v_{\mathrm{g}}\sim\Gamma/\omega_{\mathrm{a}}\ll1$, which means that under these parameters, the atom-coupled 
waveguide can be considered as a linear slow-light waveguide.      
\subsection{\label{BvsAB}$\alpha_{0}=n\pi$ and $\beta_{0}=(m-1/2)\pi$: band gap with tunable center frequency}
\begin{figure}[t]
\centering
\includegraphics[width=0.5\textwidth]{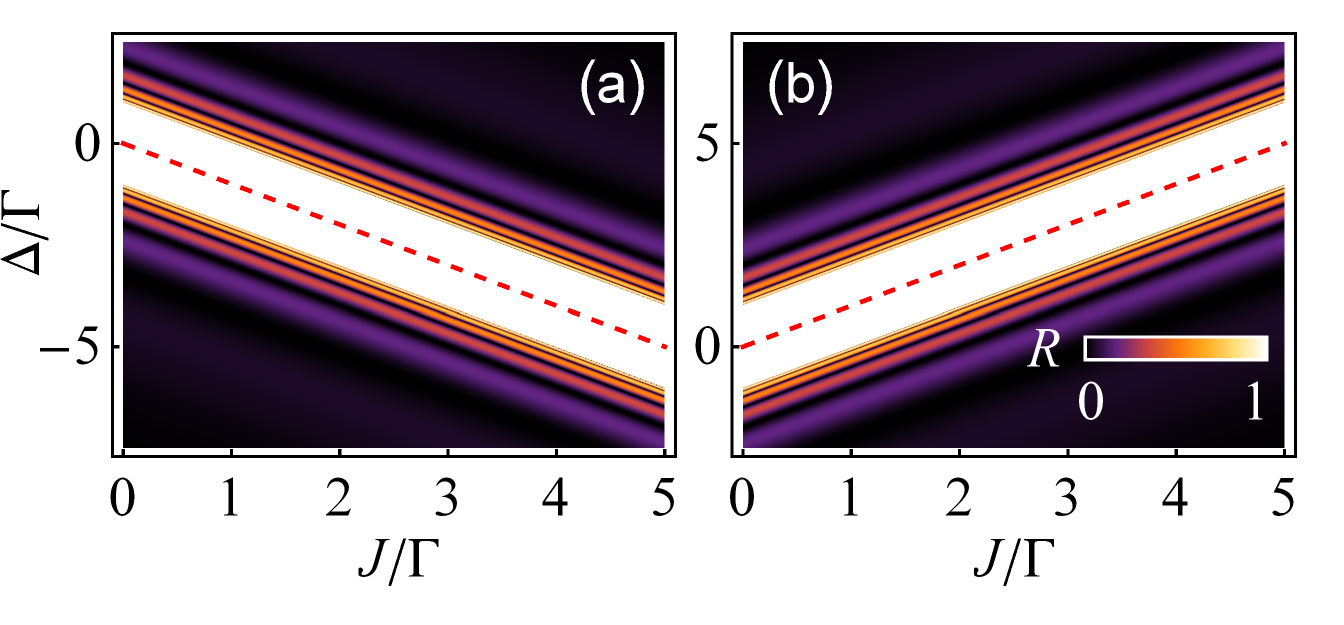}
\caption{Reflectance for a dimer chain as functions of $\Delta$ and $J$, with (a) $\alpha_{0}=\pi$, (b) $\alpha_{0}=2\pi$. 
The red dashed lines indicate the position of the center of the band gap.
Other parameters are set as $M=15$, $\beta_{0}=5\pi/2$, and $\omega_{a}=10^{4}\Gamma$.}
\label{DimerBGT2}
\end{figure}
We subsequently discuss the case of $\alpha_{0}=n\pi$ ($n\in\mathbb{N}^{+}$) and $\beta_{0}=(m-1/2)\pi$ ($m\in\mathbb{N}^{+}$, $m>n$), i.e., 
the intra(inter)-cell phase difference satisfies the Bragg (anti-Bragg) condition.
Under these conditions, Eqs.~\eqref{xidimer} and \eqref{ydimer} can be simplified as
\begin{subequations}
\begin{equation}
\zeta=\frac{\Gamma^2}{\left[\Delta-(-1)^{n} J\right]^2},
\label{xidimer3}
\end{equation}	
\begin{equation}
y=\frac{(-1)^{m+1}\Gamma}{\Delta-(-1)^{n} J}.
\label{ydimer3}
\end{equation}
\end{subequations}
Similar to Sec.~\ref{ABvsB}, with the help of the condition \eqref{BGConditionY}, we can find that the region of the band gap is
\begin{equation} 
\mp J-\Gamma<\Delta<\mp J+\Gamma,  
\label{BGRange2}
\end{equation}
indicating that the center of the band gap is located at $\Delta=\mp J$ (where ``$-$'' corresponds to $n\in\mathbb{O}^{+}$ and ``$+$'' corresponds to $n\in\mathbb{E}^{+}$), 
and the corresponding width is
\begin{equation}
W=2\Gamma.
\label{Wcase2}
\end{equation}
This result shows that the center of the band gap can be tuned by changing the intra-cell coupling strength $J$. 
As shown in Figs.~\ref{DimerBGT2}(a) (with $\alpha_0=\pi$ and $\beta_0=5\pi/2$) and \ref{DimerBGT2}(b) (with $\alpha_0=2\pi$ and $\beta_0=5\pi/2$), due to the existence of the intra-cell coupling, the center of the band gap is red-shifted or blue-shifted by $J$ in comparison with a non-interacting atomic chain [see the red dashed lines in Figs.~\ref{DimerBGT2}(a) and \ref{DimerBGT2}(b)], while the width of the band gap remains unchanged.
\begin{figure}[t]
\centering
\includegraphics[width=0.5\textwidth]{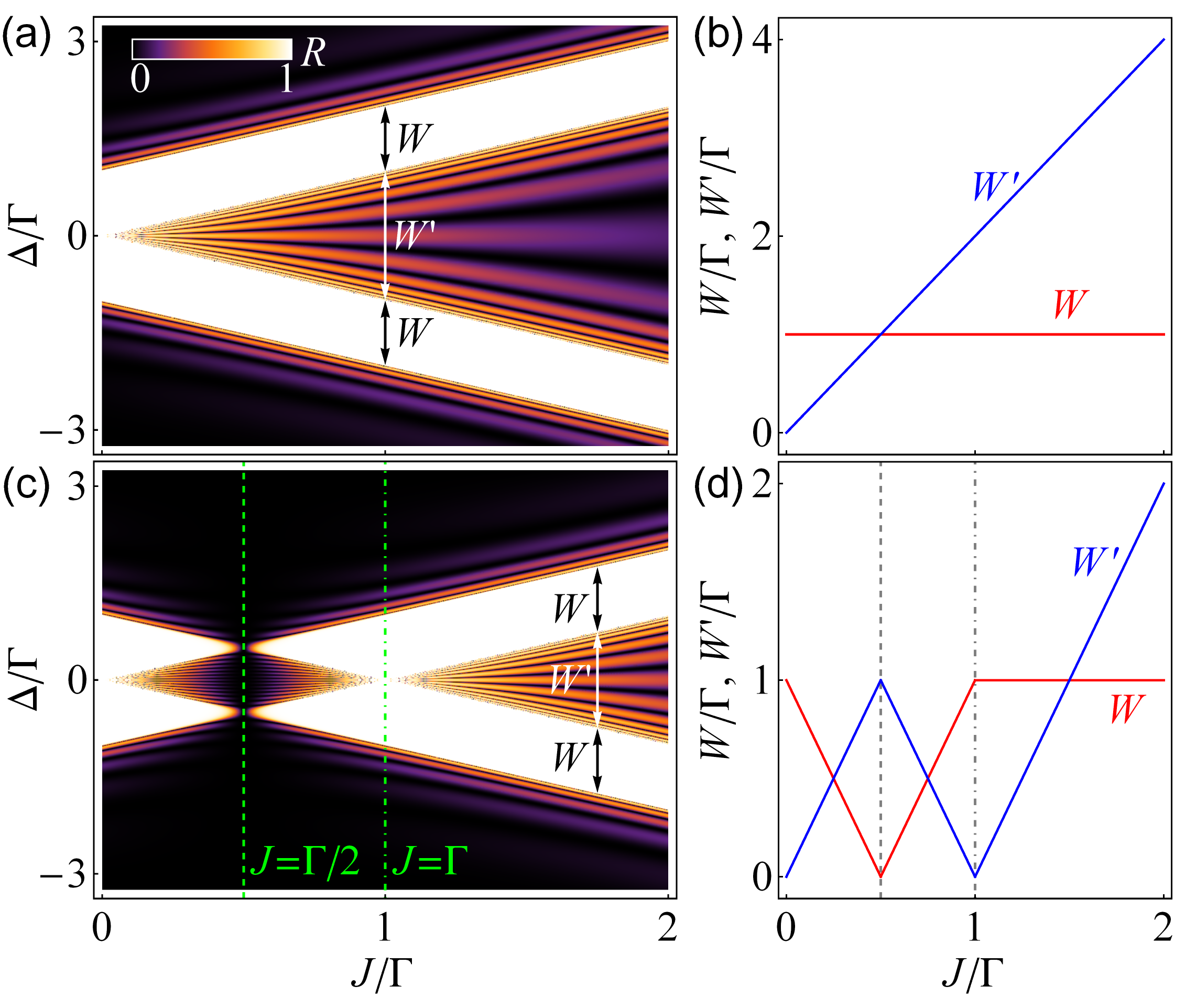}
\caption{(a) Reflectance for a dimer chain as a function of $\Delta$ and $J$. The parameters 
are set as $M=15$,  $\alpha_{0}=\pi/2$, $\beta_0=5\pi/2$, and $\omega_{\mathrm{a}}=10^{4}\Gamma$, respectively.
(b) The width $W$ of the band gap and the width $W'$ of the center passband as functions of $J$
corresponding to the case shown in panel (a).
(c) Reflectance for a dimer chain as a function of $\Delta$ and $J$, with intra-cell phase 
delay $\alpha_{0}=3\pi/2$. Other parameters are the same as those used in panel (a). 
(d) The widths $W$ and $W'$ as functions of $J$ corresponding to the case shown in panel (c).
In panels (c) and (d), the dashed lines are employed to indicate the value of $J$ corresponding to the disappearance of gaps, 
whereas the dash-dotted lines are used to indicate the disappearance of a center passband.}
\label{DimerBGT3}
\end{figure}
\subsection{\label{ABvsAB}$\alpha_{0}=(n-1/2)\pi$ and $\beta_{0}=(m-1/2)\pi$: tunable double band gaps and center passband}
\subsubsection{\label{TunableDB}Tunable band-gap structures}
Finally, we discuss the case of $\alpha_{0}=(n-1/2)\pi$ ($n\in\mathbb{N}^{+}$) and $\beta_{0}=(m-1/2)\pi$ ($m\in\mathbb{N}^{+}$, $m>n$), 
i.e., both the intra- and inter-cell phase differences satisfy the anti-Bragg condition.
For this case, Eqs.~\eqref{xidimer} and \eqref{ydimer} can be simplified as
\begin{subequations}
\begin{equation}
\zeta=\frac{\Gamma^2\left[2J-(-1)^{n}\Gamma\right]^2}{4\left\{\Delta^2-J\left[J-(-1)^{n}\Gamma\right]\right\}^2},
\label{xidimer4}
\end{equation}	
\begin{equation}
y=\frac{(-1)^{m+1}\Gamma\Delta}{\Delta^2-J\left[J-(-1)^{n}\Gamma\right]}.
\label{ydimer4}
\end{equation}
\end{subequations}

For $n\in\mathbb{O}^{+}$, different from the previous cases, there are always two regions satisfying the condition \eqref{BGConditionY}, with 
\begin{equation}
J<|\Delta|<J+\Gamma. 
\label{BGRange3a}
\end{equation}
These regions with no propagating modes will form two band gaps centered at $\Delta=\pm(J+\Gamma/2)$ with equal width 
\begin{equation}
W=\Gamma.
\label{Wcase3a}
\end{equation}
We refer to them as the right and left band gaps, respectively [see Fig.~\ref{DimerBGT3}(a)]. And the region $-J<\Delta<J$ between the two gaps forms a
passband whose width is proportional to $J$ 
\begin{equation}
W'=2J.
\label{Wpcase3a}
\end{equation}
In a word, we can obtain a symmetrical spectral structure with a passband sandwiched between two band gaps of equal width. In addition, 
the centers of the gaps, and thus the width of the passband, show a linear relationship with $J$, as shown in Fig.~\ref{DimerBGT3}(a). 
To show these more clearly, the widths $W$ and $W'$ as functions of $J$ are given in Fig.~\ref{DimerBGT3}(b).

If $n\in\mathbb{E}^{+}$, as in the case of $n\in\mathbb{O}^{+}$, two band-gap regions symmetric about $\Delta=0$ 
appear in the reflection spectra for different $J$, except for the cases of  
$J=\Gamma/2$ (where the gaps are closed) and $J=\Gamma$
(where the two gaps merge into one). But a richer modulation of the band-gap structure can be obtained by varying the 
intra-cell coupling strength $J$ [see Fig.~\ref{DimerBGT3}(c)]. According to the condition \eqref{BGConditionY}, we 
can obtain the regions of the gaps, the width $W$ of each gap, and the width $W'$ of the passband between the gaps, as 
summarized in Table \ref{WforABAB}. Correspondingly, the widths $W$ and $W'$ as functions
of $J$ are plotted in Fig.~\ref{DimerBGT3}(d). Specifically, in the region $0<J<\Gamma$, the two band gaps are centered
at $\Delta=\pm \Gamma/2$ and the width of each gap first decreases linearly (with $W=\Gamma-2J$) to zero at 
$J=\Gamma/2$ [where the gaps vanish, see the dashed lines in Figs.~\ref{DimerBGT3}(c) and \ref{DimerBGT3}(d)] and then 
increases linearly (with $W=2J-\Gamma$). Finally the two gaps merge into a single band of width $2\Gamma$ 
at $\Delta=\Gamma$ [see the dash-dotted lines in Figs.~\ref{DimerBGT3}(c) and \ref{DimerBGT3}(d)]. 
Accordingly, the width of the center passband first increases linearly (with $W'=2J$) to a local maximum 
when $J\to\Gamma/2$ (note that at $J=\Gamma/2$, the center passband cannot be defined because 
the gaps are closed) and then decreases linearly [with $W'=2(\Gamma-J)$] to zero at 
$\Delta=\Gamma$. In the region $J>\Gamma$, the gap splits again into two, 
centered at $\Delta=\pm(J-\Gamma/2)$ and of equal width $J$. And the width of the corresponding center passband increases 
linearly [with $W'=2(J-\Gamma)$].    
\begin{table}[t]
\renewcommand{\arraystretch}{1.5}
\centering\caption{Summary of the regions of the gaps, the width $W$ of the two gaps, and the width $W'$ of the center passband 
between the gaps for different ranges of $J$ for the case of $n\in\mathbb{E}^{+}$ discussed in Sec.~\ref{ABvsAB}.}
\label{table}
\begin{ruledtabular}
\begin{tabular}{cccccccc}
&Range of $J$
&Gap regions
&$W$
&$W'$
\\
\hline
&$0<J<\Gamma/2$
&$ J<|\Delta|<\Gamma-J$
&$\Gamma-2J$
&$2J$
\\
&$\Gamma/2<J<\Gamma$
&$\Gamma-J<|\Delta|<J$
&$2J-\Gamma$
&$2(\Gamma-J)$
\\
&$J>\Gamma$
&$J-\Gamma<|\Delta|<J$
&$\Gamma$
&$2(J-\Gamma)$
\\    
\end{tabular}
\end{ruledtabular}
\label{WforABAB}
\end{table}
\subsubsection{\label{EngineeringPB}Center passband engineering}
\begin{figure}[t]
\centering
\includegraphics[width=0.5\textwidth]{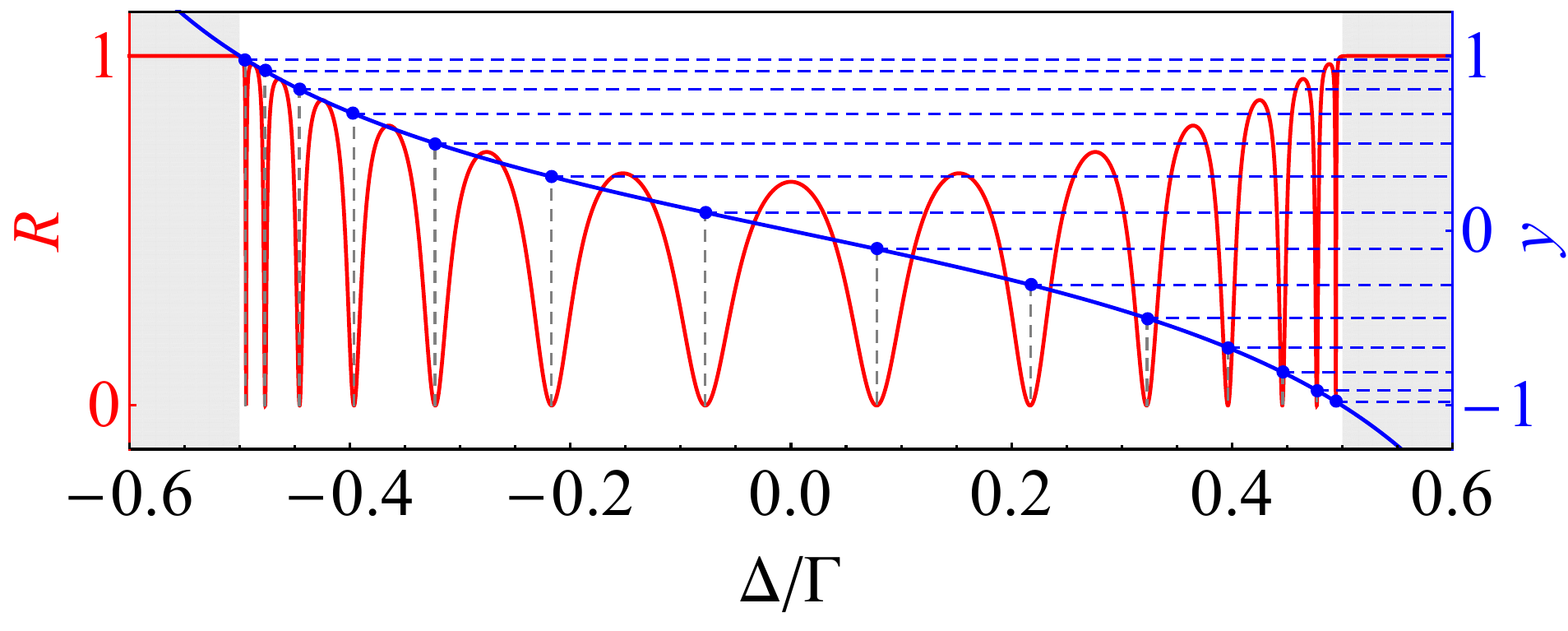}
\caption{The reflection spectrum and the function $y(\Delta)$ in the region of the central passband. The intra-cell coupling strength 
is set as $J=3\Gamma/2$, and the remaining parameters are the same as those utilized in Fig.~\ref{DimerBGT3}(c).
The horizontal dashed lines are given by $y=y_{l}=\cos(l\pi/M)$ (from top to bottom: $l=1,2,\cdots,M-1$), and the values of $\Delta$ 
indicated by the vertical dashed lines are the solutions to the equation $y(\Delta)=y_{l}$, which is consistent with the points of $R=0$. 
These solutions correspond to the frequencies of the propagating modes in the passband.
The band gaps are represented by the shaded region (with $|y(\Delta)|>1$).}
\label{CMIPB}
\end{figure}
\begin{figure}[t]
\centering
\includegraphics[width=0.5\textwidth]{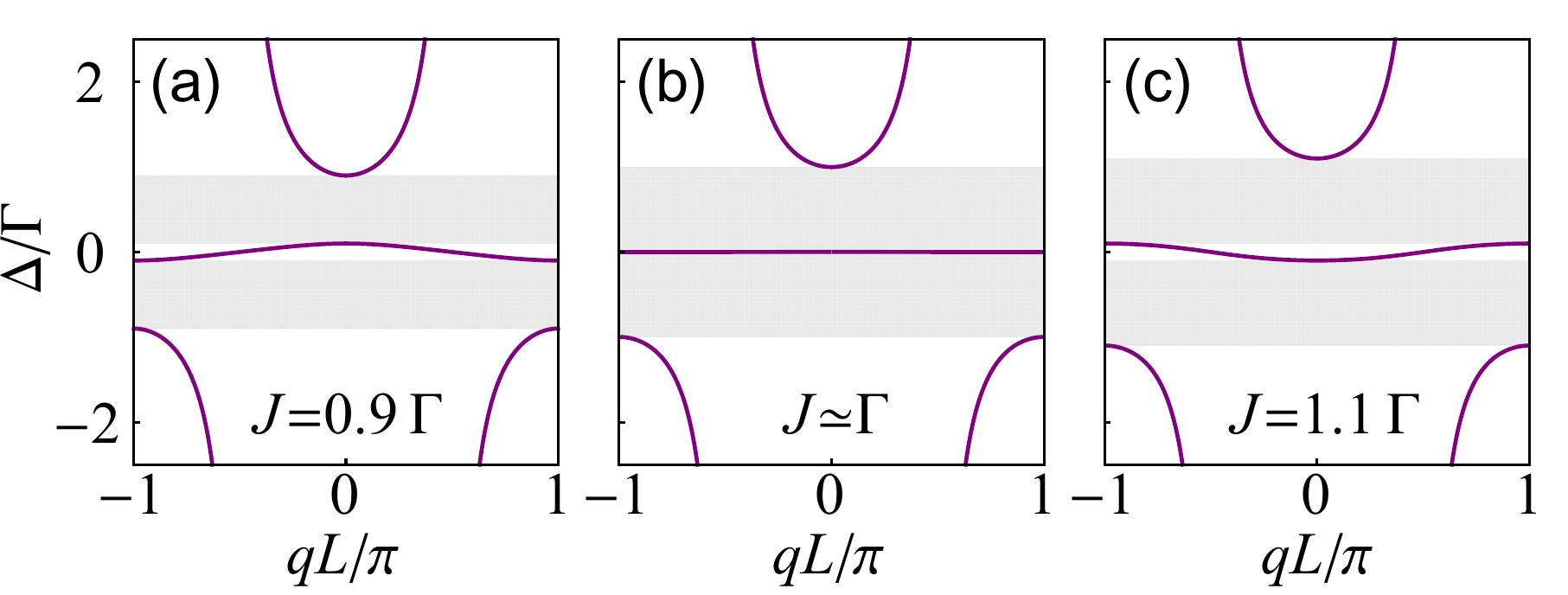}
\caption{Dispersion relations for a dimer chain for various values of $J$ in the vicinity of $J=\Gamma$. 
The parameters are the same as those used in Fig.~\ref{DimerBGT3}(c).}
\label{DimerDisRela3}
\end{figure}
Compared to the previous subsections, the most significant feature of the situation now discussed is the 
possibility of realizing a highly engineered center passband. As we have discussed earlier, its width is easily tunable 
by the intra-cell coupling strength $J$. Here we will further discuss its other properties.    
As discussed in Sec.~\ref{FormationBG}, the solutions to Eq.~\eqref{ys} can give the 
zero reflection points in the reflection spectrum. For present case, as shown in Fig.~\ref{CMIPB}, 
the center branch of $y(\Delta)$ can be used to fix the $M-1$ zero reflection points within the passband, 
which correspond to the propagating modes permitted by this band. According to Eqs.~\eqref{ys} 
and \eqref{DispersionR}, one can find that the wave numbers of these propagating modes are $q=\pm l\pi/(ML)$ ($l=1,2,\cdots,M-1$).

If the frequencies of photons are concentrated within the center passband, the atom-perturbed system can be regarded
as a nonlinear waveguide with finite bandwidth, with the group velocity being considerably less than that of the 
propagating photons in a bare waveguide. By changing $J$, the band width and thus the dispersion relation of the center 
passband can also be easily engineered. To illustrate, we consider the case of $n\in\mathbb{E}^{+}$ and start with a 
coupling strength that falls within the range of $\Gamma/2 < J < \Gamma$ [see Fig.~\ref{DimerDisRela3}(a)]. 
Subsequently, the value of $J$ is increased until it reaches $J\simeq \Gamma$, during which the bandwidth of the 
center passband is compressed and ultimately reduces to zero. Consequently, the center passband becomes a flat 
band, as illustrated in Fig.~\ref{DimerDisRela3}(b). It is evident that, in this situation, the density of the traveling modes 
becomes exceedingly high, while the group velocity for photons in this band is close to zero. As a result, it is
possible to slow or even stop a single-photon pulse at will in this setup by dynamically controlling $J$. 
If the coupling strength $J$ is precisely equal to $\Gamma$, the center passband will disappear, 
as the two gaps will merge into a single gap of width $2\Gamma$. As the value of $J$ is increased to satisfy 
$J>\Gamma$, the group velocity of the center passband undergoes a change in sign, as illustrated in 
Fig.~\ref{DimerDisRela3}(c). This indicates that the sign of the group velocity (i.e., the direction of wave-packet 
propagation) for the passband can also be controlled by selecting an appropriate $J$. We note that similar 
modulations can also be achieved in a wQED structure containing a resonator array \cite{Yanik-PRL2004} or a  chain of superconducting qubits \cite{Shen-PRB2007} through the appropriate adjustment of the detuning between the emitters in a cell.

Similar to the second case discussed in Sec.~\ref{ABvsB}, 
when $n\in\mathbb{E}^{+}$ and $J=\Gamma/2$ [where the gaps are closed, indicated by the dashed lines in 
Figs.~\ref{DimerBGT3}(c) and \ref{DimerBGT3}(d)], the atom-coupled waveguide can be utilized as 
a linear waveguide when the modes in the vicinity of $\Delta=\pm \Gamma/2$ are of interest. When
$\alpha_{0}=3\pi/2$, $\beta_0=5\pi/2$ [the parameters used in Fig.~\ref{DimerBGT3}(c)], 
the corresponding dispersion relations in the linear regions can be approximated as 
$|\Delta-\Gamma/2|=\tilde{v}_{\mathrm{g}}|q|$ and $|\Delta+\Gamma/2|=\tilde{v}_{\mathrm{g}}|q\pm\pi/L|$, respectively. 
Here $\tilde{v}_{\mathrm{g}}=\Gamma L/2$ is the group velocity at $\Delta=\pm \Gamma/2$ of the perturbed waveguide. 
We also have $\tilde{v}_{\mathrm{g}}/v_{\mathrm{g}}\sim\Gamma/\omega_{\mathrm{a}}\ll1$, which indicates the 
slow-light effect. 
\begin{figure*}[t]
\centering
\includegraphics[width=\textwidth]{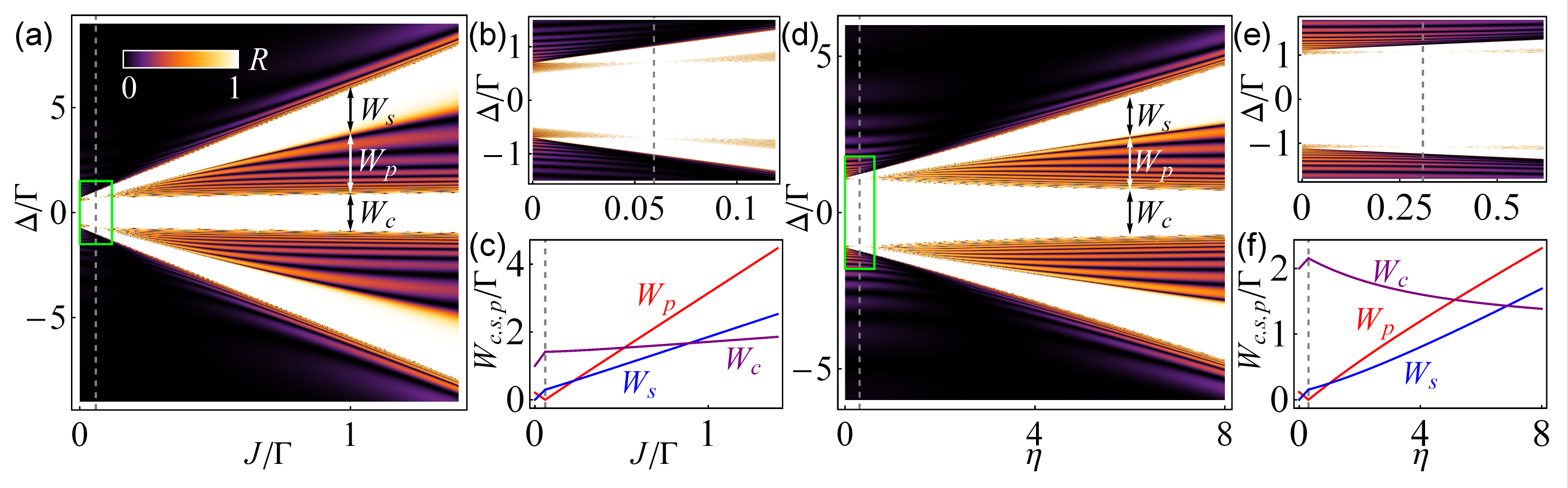}
\caption{(a) Reflectance for a tetramer chain as a function of $\Delta$ and $J$ for fixed $\eta=5$.
(b) The details in the box in panel (a). 
(c) The widths $W_{c}$, $W_{s}$, and $W_{p}$ as functions of $J$ for fixed $\eta=5$. 
(d) Reflectance for a tetramer chain as a function of $\Delta$ and $\eta$ for fixed $J=\Gamma/2$.
(e) The details in the box in panel (d). 
(f) The widths $W_{c}$, $W_{s}$, and $W_{p}$ as functions of $\eta$ for fixed $J=\Gamma/2$. 
The gray dashed lines in (a)-(f) are used to mark the positions of $f(\tilde{J},\eta)=0$, at which the three gaps converge to form a 
single one. Other parameters are set as $M=15$, 
$\alpha_{0}=\pi/2$, $\beta_{0}=2\pi$ and $\omega_{\mathrm{a}}=10^{4}\Gamma$.}
\label{TetramerBGT}
\end{figure*}
\section{\label{TetramerChain}Tunable photonic band gaps with a polymer chain}
The discussions in last section can be generalized to the case of a polymer chain, wherein each unit cell contains a greater number of atoms than two. Naturally, as the number of atoms in a cell increases, multi-band-gap structure
can be realized. Furthermore, the design of intra-cell interactions enables the implementation of more sophisticated 
band-gap engineering. In this section, we utilize the case of a tetramer ($N=4$) array coupled to a waveguide, as shown schematically in Fig.~\ref{Sketch}(b), as an example
to discuss these properties. The coupling strengths between neighboring atoms in a cell
are set as  $J_{1}=J_{3}=J$ and $J_{2}=\eta J$, where the dimensionless parameter $\eta$ is 
introduced to enhance the tunability of the system. 
The phase factors are set as $\alpha_{0}=\pi/2$ and $\beta_{0}=2\pi$. 
Correspondingly, the function $y(\Delta)$ under the Markovian approximation takes the form 
\begin{equation}
y(\Delta)=\frac{\Delta^{4}-\mathcal{A}\Delta^{2}+\mathcal{B}}{\Delta^{4}-\mathcal{C}\Delta^{2}+\mathcal{D}},
\label{ploymery}
\end{equation}
where
\begin{equation}
\begin{aligned}
&\mathcal{A}=2\Gamma^{2}+\left(\eta+2\right)\Gamma J+\left(\eta^{2}+2\right)J^{2},
\\
&\mathcal{B}=\left(\Gamma^{2}+J^{2}\right)\left[\left(\eta+2\right)\Gamma J+J^2\right]
+\frac{1}{2} \Gamma^{2}\left[\Gamma^{2}+\left(\eta+2\right)^{2}J^{2}\right], 
\\
&\mathcal{C}=\mathcal{A}-2\Gamma^{2},~~\mathcal{D}=\left[\left(\Gamma+J\right)^{2}+\eta\Gamma J\right]J^{2}.
\nonumber
\end{aligned}
\label{abcd}
\end{equation}

Substituting Eq.~\eqref{ploymery} into the band-gap condition \eqref{BGConditionY}, 
we can obtain the band-gap regions for different values of $J$ and $\eta$. To
classify the solutions to the inequality \eqref{BGConditionY}, 
one can define the following discriminant function 
\begin{equation}
f\left(\tilde{J},\eta\right)=3\tilde{J}^{2}\eta^{2}+2\left(4\tilde{J}^2+\tilde{J}\right)\eta-1, 
\end{equation}
with $\tilde{J}=J/\Gamma $. For 
$f(\tilde{J},\eta)\neq 0$, the reflection spectra indicate that a triple-band-gap structure, 
comprising a center gap and two side gaps of equal width, is distributed symmetrically about $\Delta=0$, 
as shown in Figs.~\ref{TetramerBGT}(a) (for fixed $\eta=5$ and varying $J$) 
and \ref{TetramerBGT}(d) (for fixed $J=\Gamma/2$ and varying $\eta$). 
The details of these spectra around the regions satisfying $f(\tilde{J},\eta)= 0$ are provided
in Figs.~\ref{TetramerBGT}(b) and \ref{TetramerBGT}(e).
Furthermore, both the centers and the widths of the two side gaps and the two passbands 
between the gaps are highly tunable by varying the values of $J$ and $\eta$.  
The regions of the gaps, the width $W_{c}$ of the center gap, the width $W_{s}$ of the two side gaps, 
and the width $W_p$ of the two passbands between the gaps are 
summarized in Table \ref{WcWsWp}. The quantities $\delta$ and $\tilde{\delta}_{1,2}$ in the table take the form
\begin{subequations} 
\begin{equation}
\delta=\frac{1}{2}\left[\Gamma+\left(\eta+2\right)J\right],
\label{delta}
\end{equation}
\begin{equation}
\tilde{\delta}_{1,2}=\frac{1}{2}\left[\pm\eta J+\sqrt{\left(\Gamma+2J\right)^{2}+\left(\Gamma+\eta J\right)^{2}}\right].
\label{delta1}
\end{equation}
\end{subequations}
It can be demonstrated that the relations $0<\tilde{\delta}_2<\tilde{\delta}_1$, $0<\delta<\tilde{\delta}_1$, and
\begin{equation}
\left.\delta-\tilde{\delta}_2\left\{\begin{array}{ll}>0,&f\big(\tilde{J},\eta\big)>0\\[2ex]=0,
&f\big(\tilde{J},\eta\big)=0\\[2ex]<0,
&f\big(\tilde{J},\eta\big)<0\end{array}\right.\right.
\nonumber
\label{deltajudge}
\end{equation}
are satisfied. In accordance with the reflection spectra illustrated in Figs.~\ref{TetramerBGT}(a) and \ref{TetramerBGT}(d), the widths $W_c$, $W_s$, and $W_p$ are plotted as functions of $J$ (for fixed $\eta$) and of $\eta$ (for fixed $J$)  in Figs.~\ref{TetramerBGT}(c) and \ref{TetramerBGT}(f), respectively.
These results show that a tetramer chain can provide greater flexibility in band-gap engineering.
\begin{table}[b]
\renewcommand{\arraystretch}{1.75}
\centering\caption{Summary of the regions of the gaps, the width $W_{c}$ of the center gap, the width $W_{s}$ of the two side gaps, 
and the width $W_p$ of the two passbands between the gaps for the setup discussed in Sec.~\ref{TetramerChain} under condition
$f\big(\tilde{J},\eta\big)<0$ or $f\big(\tilde{J},\eta\big)>0$.}
\begin{ruledtabular}
\begin{tabular}{cccccccc}
&Condition
&Gap regions
&$W_{c}$
&$W_{s}$
&$W_{p}$
\\
\hline
&$f\big(\tilde{J},\eta\big)<0$
&$|\Delta|<\delta\cup\tilde{\delta}_{2}<|\Delta|<\tilde{\delta}_{1}$
&$2\delta$
&$\tilde{\delta}_{1}-\tilde{\delta}_{2}$
&$\tilde{\delta}_{2}-\delta$
\\
&$f\big(\tilde{J},\eta\big)>0$
&$|\Delta|<\tilde{\delta}_{2}\cup\delta<|\Delta|<\tilde{\delta}_{1}$
&$2\tilde{\delta}_{2}$
&$\tilde{\delta}_{1}-\delta$
&$\delta-\tilde{\delta}_{2}$
\\    
\end{tabular}
\end{ruledtabular}
\label{WcWsWp}
\end{table}

\begin{figure}[t]
\centering
\includegraphics[width=0.5\textwidth]{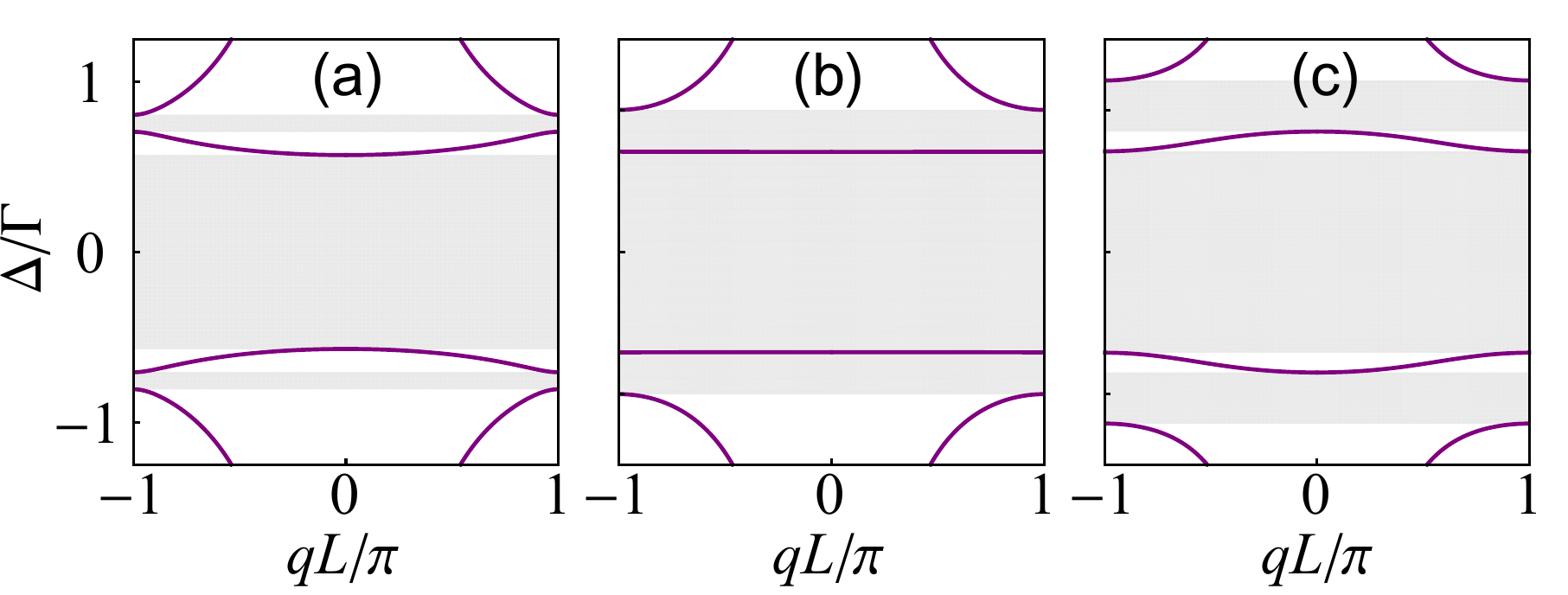}
\caption{Dispersion relations for a tetramer chain, with fixed $\eta=5$ and varying $J$: (a) $J=0.02\Gamma$ [satisfy $f(\tilde{J},\eta)<0$], (b) $J=0.059\Gamma$ [satisfy $f(\tilde{J},\eta)\simeq 0$] and (c) $J=0.1\Gamma$ [satisfy $f(\tilde{J},\eta)> 0$].
Other parameters are set as $\alpha_{0}=\pi/2$, $\beta_{0}=2\pi$, and $\omega_{\mathrm{a}}=10^{4}\Gamma$.}
\label{TetramerDisRela}
\end{figure}

In the case of $f(\tilde{J},\eta)=0$, the three band gaps merge into a single one centered about $\Delta=0$, with width $2\tilde{\delta}_1$, as 
illustrated by the dashed lines in Figs.~\ref{TetramerBGT}(a)-\ref{TetramerBGT}(f). Importantly, if
we select an appropriate path in the parameter space spanned by $(\eta,J)$ [i.e., from the area satisfying $|f(\tilde{J},\eta)|>0$
to a point on the curve $f(\tilde{J},\eta)=0$], the two passbands
between the gaps can be easily compressed and ultimately becomes a flat band when $f(\tilde{J},\eta)\to 0$. 
This property is illustrated in Figs.~\ref{TetramerDisRela}(a)-\ref{TetramerDisRela}(c), in which $\eta$ is held constant 
while $J$ is allowed to vary.  Furthermore, the values of $\eta$ and $J$ employed in these figures have been selected to 
satisfy the conditions $f(\tilde{J},\eta)<0$, $f(\tilde{J},\eta)\simeq 0$, and  $f(\tilde{J},\eta)> 0$, respectively.
Thus, like the case discussed in Sec.~\ref{EngineeringPB}, it is possible to slow or even stop a single-photon pulse 
by modulating the relevant parameters. However, this time we are able to control the propagation of photon pulses in two 
distinct passbands simultaneously. Finally, it can be seen from Figs.~\ref{TetramerDisRela}(a) and \ref{TetramerDisRela}(c) 
that the group velocities of the two passbands between the gaps can also be designed
by selecting suitable values of the parameters $\eta$ and $J$. 
\section{\label{Conclusions}Conclusions}
In this paper, we have analyzed the generation and engineering of photonic band gaps in wQED systems containing multiple atom-polymers.
We first consider the configuration of a dimer array coupled to a waveguide with the intra- and inter-cell phase 
delays satisfying either the Bragg or anti-Bragg condition. It was found that, with the exception of both the 
intra- and inter-cell phase delays satisfying the Bragg condition, the photonic band-gap structures can be 
observed in the spectra for the remaining three cases. More importantly, the band-gap structures in these 
cases can be manipulated in a variety of ways (e.g., the center and  the width of the gaps, as well as the 
dispersion relation of the passbands) by adjusting the intra-cell coupling strength. 
Furthermore, modulation of the dispersion relation and bandwidth provides an effective way to manipulate the 
propagating modes in the waveguide, leading to some interesting phenomena such as slowing or even 
stopping a single-photon pulse. Finally, the results for a dimer array can be generalized to the case of a 
greater number of atoms in each unit cell. This enables the realization of a tunable multi-gap structure and the 
implementation of more sophisticated band-gap engineering. 

In our proposal, the key point is the use of tunable couplings between atoms to realize band-gap engineering. 
This can be easily achieved in micro- and nano-quantum systems such as superconducting quantum circuits 
\cite{Gu-PhysReports2017}, where superconducting qubits or superconducting quantum interference devices (SQUIDs) are usually used as tunable coupling elements 
\cite{Hime-Science2006,Niskanen-Science2007,Baust-PRB2015,Zhu-PRA2019}. Thus, in summary, we 
present an experimentally feasible proposal to produce tunable band-gap structures in 1D wQED setups, 
which may provide novel ways to control single-photon propagation. Our proposal may have potential 
applications in long-distance quantum networking and quantum communication.
\begin{acknowledgments}
This work was supported by the National Natural Science Foundation of China (NSFC) under Grants No. 61871333.
\end{acknowledgments}

\appendix
\section{\label{DerivationEQ23} Derivation of Eqs.~\eqref{Transmission} and \eqref{Reflection}}
According to Eq.~\eqref{decomposeTM}, one can obtain 
\begin{subequations}
\begin{equation}
\left[\tilde{\mathcal{T}}^{M}\right]_{11}=\tilde{\mathcal{T}}_{11}U_{M-1}(y)-U_{M-2}(y),
\label{TM11}
\end{equation}	
\begin{equation}
\left[\tilde{\mathcal{T}}^{M}\right]_{21}=\tilde{\mathcal{T}}_{21}U_{M-1}(y).
\label{TM21}
\end{equation}
\end{subequations}
Substituting these results into Eq.~\eqref{TotalAmp} and taking the square of the modulus yields the transmission and reflection 
coefficients 
\begin{subequations}
\begin{equation}
T=|t|^2=\frac{1}{\left|\tilde{\mathcal{T}}_{11}U_{M-1}(y)-U_{M-2}(y)\right|^2},
\label{abst1}
\end{equation}	
\begin{equation}
R=|r|^{2}=\frac{\left|\tilde{\mathcal{T}}_{21}\right|^{2}U_{M-1}^2(y)}{\left|\tilde{\mathcal{T}}_{11}U_{M-1}(y)-U_{M-2}(y)\right|^{2}}.
\label{absr1}
\end{equation}
\end{subequations}
Note that the elements of the transfer matrix $\tilde{\mathcal{T}}$ satisfy the relation 
$\mathrm{Det}(\tilde{\mathcal{T}})=1$, $\tilde{\mathcal{T}}_{11}^{*}=\tilde{\mathcal{T}}_{22}$ and 
$\tilde{\mathcal{T}}_{12}^{*}=\tilde{\mathcal{T}}_{21}$ [see Eq.~\eqref{Tmatrix2}]. Thus we have 
$|\tilde{\mathcal{T}}_{11}|^2-|\tilde{\mathcal{T}}_{21}|^2=1$ and 
$\tilde{\mathcal{T}}_{11}+\tilde{\mathcal{T}}_{11}^{*}=\tilde{\mathcal{T}}_{11}+\tilde{\mathcal{T}}_{22}=\mathrm{Tr}(\tilde{\mathcal{T}})=2y$. 
Using these results and the identity
\begin{equation}
U^2_{M-1}(y)+U^2_{M-2}(y)-2yU_{M-1}(y)U_{M-2}(y)=1,
\label{idetityn1}
\end{equation}
the dominator of $T$ and $R$ can be further simplified as
\begin{eqnarray}
&&\left|\tilde{\mathcal{T}}_{11}U_{M-1}(y)-U_{M-2}(y)\right|^2
\nonumber
\\
&=&\left|\tilde{\mathcal{T}}_{11}\right|^{2}U^2_{M-1}(y)+U^2_{M-2}(y)-(\tilde{\mathcal{T}}_{11}+\tilde{\mathcal{T}}_{11}^{*})U_{M-1}(y)U_{M-2}(y)
\nonumber
\\
&=&\left(1+\left|\tilde{\mathcal{T}}_{21}\right|^{2}\right)U^2_{M-1}(y)+U^2_{M-2}(y)-2yU_{M-1}(y)U_{M-2}(y)
\nonumber
\\
&=&1+\left|\tilde{\mathcal{T}}_{21}\right|^{2}U^2_{M-1}(y).
\label{TRdominator}
\end{eqnarray}
Finally, by defining the quantity $\zeta=|\tilde{\mathcal{T}}_{21}|^{2}=|\tilde{\mathcal{T}}_{12}|^{2}$, one can obtain the transmittance and reflectance 
described by Eqs.~\eqref{Transmission} and \eqref{Reflection}. 
\section{\label{MarkovCond}The Markovian conditions}
In this paper, we focus on the regime in which the detuning-dependent phase delays defined in Eq.~\eqref{alphabeta}
can be approximated as $\alpha\simeq\alpha_0$ and $\beta\simeq\beta_0$. 
Here we will present the condition that allows us to make this approximation. 
The accumulated phase factor between two ends of the atomic chain 
under consideration in this paper is 
$(M-1)\beta=(1+\Delta/\omega_{\mathrm{a}})(M-1)\beta_0$. Obviously, for a not so small $M$, if the 
term caused by the detuning effects $(M-1)\beta_0\Delta/\omega_{\mathrm{a}}\simeq M\beta_0\Delta/\omega_{\mathrm{a}}\ll1$, it can be safely neglected. Note that for the band-gap cases (the cases in Secs.~\ref{ABvsB}-\ref{ABvsAB} and \ref{TetramerChain}), the detuning range we are interested in is approximately $\Delta\sim\Gamma$, 
and the order of magnitude of $\beta_0$ we care about is $\beta_0\sim 1$, so we end up with 
\begin{equation}
M\ll\frac{\omega_{\mathrm{a}}}{\Gamma},
\label{MarkovCondition1} 
\end{equation}
which is easily satisfied under typical parameters for wQED systems \cite{Sheremet-RMP2023}. And for the case discussed in Sec.~\ref{BvsB}), we have $\Delta\sim M\Gamma$, the Markovian condition becomes
\begin{equation}
M\ll\sqrt{\frac{\omega_{\mathrm{a}}}{\Gamma}}.
\label{MarkovCondition2} 
\end{equation}
Furthermore, under these conditions the detuning-dependent term corresponding to any other pair of coupling points can also be ignored,
since the distance between them is smaller than the length of the atomic chain.
In a word, once the condition \eqref{MarkovCondition1} (valid for the cases discussed in 
Secs.~\ref{ABvsB}-\ref{ABvsAB} and \ref{TetramerChain}) or \eqref{MarkovCondition2}
(valid for the case discussed in Sec.~\ref{BvsB}) is satisfied, we can safely let $\alpha\simeq\alpha_0$ and $\beta\simeq\beta_0$.
\section{\label{DerivationDR}Derivation of dispersion relation}
Consider an infinite periotic polymer chain, we write the photonic wave functions to the immediate left and right of the $m$th unit cell as
\begin{subequations}
\begin{equation}
u^{(k)}_{m}\left(x\right)=t_{m-1,N}e^{ikx}+r_{m,1}e^{-ikx},
\label{um}
\end{equation}
\begin{equation}
u^{(k)}_{m+1}\left(x\right)=t_{m,N}e^{ikx}+r_{m+1,1}e^{-ikx}.
\label{um1}
\end{equation}
\end{subequations}
According to Bloch's theorem, we have $u_{m+1,k}\left(x+L\right)=u_{m,k}\left(x\right)e^{iqL}$, where $q$ is the Bloch wave number, which
leads to the following relations
\begin{equation}
t_{m,N}=t_{m-1,N}e^{i\left(qL-\beta\right)},~~ r_{m+1,1}=r_{m,1}e^{i\left(qL+\beta\right)}.
\label{trBloch}
\end{equation}
Substituting Eqs.~\eqref{Tmatrix2} and \eqref{trBloch} into Eq.~\eqref{TConnect1} yields the following linear equations for  
$t_{m-1,N}$ and $r_{m,1}$:
\begin{equation}
\renewcommand{\arraystretch}{1.5}
\left(\mathscr{B}^{-1}_{m}\mathcal{T}\mathscr{B}_{m-1}-e^{-iqL}\mathcal{I}\right)
\begin{pmatrix}
t_{m-1,N}
\\
r_{m,1}
\end{pmatrix}
=0.
\label{linearequations}
\end{equation}
The non-zero solutions for $t_{m-1,N}$ and $r_{m,1} $ requires the determinant of the coefficient matrix to be zero, resulting in 
the following dispersion relation
\begin{equation}
\cos\left(qL\right)=\mathrm{Re}\left(\frac{1}{\tilde{t}}e^{-i\beta}\right)
=\frac{1}{2}\mathrm{Tr}\left(\tilde{\mathcal{T}}\right)=y\left(\Delta\right).
\label{indirectdispersionrelation}
\end{equation}
\section{\label{CollectiveModes}Expressions of scattering amplitudes in terms of collective modes}
By starting from the transport equations \eqref{EoM1}-\eqref{EoM3} and performing some algebra,
we can obtain the scattering amplitudes of the atomic chain \cite{Peng-PRA2023}
\begin{subequations}
\begin{equation}
t=1-i\mathbf{W}^{\dagger}(\Delta\mathbb{I}-\mathbf{H})^{-1}\mathbf{W},
\label{tcollective}
\end{equation}
\begin{equation}
r=-i\mathbf{W}^{\top}(\Delta\mathbb{I}-\mathbf{H})^{-1}\mathbf{W}.
\label{rcollective}
\end{equation}
\end{subequations}
Here $\mathbb{I}$ is the $N'$-dimensional identity matrix. $N'=MN$ is the total number of atoms. $\mathbf{W}$ takes the form 
\begin{equation}
\mathbf{W}=\sqrt{\frac{\Gamma}{2}}\left(e^{i\phi_1},e^{i\phi_2},\cdots,e^{i\phi_{N'}}\right)^\mathrm{\top}.
\label{Wmatrix}
\end{equation}
$\mathbf{H}$ is the effective non-Hermitian Hamilton matrix of the atomic array, with elements 
\begin{equation}
\mathcal{H}_{nn'}=\mathcal{J}_{n}\delta_{n,n'-1}+\mathcal{J}_{n'}\delta_{n-1,n'}-i\frac{\Gamma}{2}e^{i\left|\phi_{n}-\phi_{n'}\right|},
\label{EffH}
\end{equation}
with $n,n'=1,2,\cdots,N'$. Here the coupling strength $\mathcal{J}_n$ between neighboring atoms 
is defined as $\mathcal{J}_{mN+s}=J_{s}$ and $\mathcal{J}_{mN}=0$ ($m=0,1,2,\cdots,M-1$ and $s=1,2,\cdots,N-1$).
The first two terms of Eq.~\eqref{EffH} describe the direct atom-atom coupling, and the last term
summarizes the atom-waveguide decay (for $n=n'$) as well as the coherent and dissipative atom-atom interactions mediated by the 
waveguide modes  (for $n\neq n'$). 

To better understand the physical properties of the scattering process, we rewrite the scattering amplitudes in terms of collective modes of the 
atomic chain:
\begin{subequations}
\begin{equation}
t=1-i\sum_{j=1}^{N'}\frac{\left(\mathbf{W}^{\dagger}\mathbf{U}_{j}^{\mathscr{R}}\right)
\left(\mathbf{U}_{j}^{\mathscr{L}^{\dagger}}\mathbf{W}\right)}{\Delta-\lambda_{j}},
\label{tDecompose1}
\end{equation}
\begin{equation}
r =-i\sum_{j=1}^{N'}\frac{\left(\mathbf{W}^{\top}{\mathbf{U}^{\mathscr{R}}_{j}}\right)
\left({\mathbf{U}^{\mathscr{L}}_{j}}^\dagger\mathbf{W}\right)}{\Delta-\lambda_j},
\label{rDecompose1}
\end{equation}	
\end{subequations}
where $\mathbf{U}^{\mathscr{R}}_{j}$ and $\mathbf{U}^{\mathscr{L}}_{j}$ are the right and left eigenvectors of the non-Hermitian Hamilton matrix 
$\mathbf{H}$, and $\lambda_j$ and $\lambda^{*}_j$ are the corresponding complex eigenvalues, satisfying 
$\mathbf{H}\mathbf{U}^{\mathscr{R}}_{j}=\lambda_{j}\mathbf{U}^{\mathscr{R}}_{j}$, 
$\mathbf{H}^{\dag}\mathbf{U}^{\mathscr{L}}_{j}=\lambda^{*}_{j}\mathbf{U}^{\mathscr{L}}_{j}$,
and ${\mathbf{U}^{\mathscr{L}}_{j}}^\dag\mathbf{U}^{\mathscr{R}}_{j'}={\mathbf{U}^{\mathscr{R}}_{j}}^\dag\mathbf{U}^{\mathscr{L}}_{j'}=\delta_{jj'}$ \cite{Brody-JPA2013}. 

In the Markovian regime we are interested in, the phase accumulation effect caused by the detuned photons can be ignored, 
thus the quantities $\mathbf{W}$, the non-Hermitian Hamiltonian $\mathbf{H}$, and the corresponding eigenvalue $\lambda_{j}$ 
and eigenvector $\mathbf{U}_{j}^{\mathscr{L, R}}$ are independent of the detuning $\Delta$ of photons. 
The transmission and reflection amplitudes [Eqs.~\eqref{tDecompose1} and \eqref{rDecompose1}] can be further written as
\begin{subequations}
\begin{equation}
t=1+\sum_{j=1}^{N'}{\frac{c_j}{\Delta-\tilde{\Delta}_j+i\frac{\tilde{\Gamma}_j}{2}}},
\label{tDecompose2}
\end{equation}
\begin{equation}
r=\sum_{j=1}^{N'}{\frac{d_j}{\Delta-\tilde{\Delta}_j+i\frac{\tilde{\Gamma}_j}{2}}}.
\label{rDecompose2}
\end{equation}	
\end{subequations}
These results show superpositions of several Lorentzian-type amplitudes contributed by the collective excitations. 
$\tilde{\Delta}_j=\mathrm{Re}(\lambda_j)|_{\Delta=0}$ is the detunning between the $j$th collective mode and the atoms, and 
$\tilde{\Gamma}_{j}=-2\mathrm{Im}(\lambda_{j})|_{\Delta=0}$ is the effective decay of the $j$th collective mode. 
$c_{j}=-i[(\mathbf{W}^{\dagger}\mathbf{U}_{j}^\mathscr{R})(\mathbf{U}_{j}^{\mathscr{L}^{\dagger}}\mathbf{W})]|_{\Delta=0}$ 
and $d_{j}=-i[(\mathbf{W}^{\top}\mathbf{U}_{j}^\mathscr{R})(\mathbf{U}_{j}^{\mathscr{L}\dagger}\mathbf{W})]|_{\Delta=0}$
determine the weight of each Lorentzian component.  
%
%
\bibliography{MS-WMS-July-31-2024}

\providecommand{\noopsort}[1]{}\providecommand{\singleletter}[1]{#1}%
\begin{thebibliography}{65}%
\makeatletter
\providecommand \@ifxundefined [1]{%
 \@ifx{#1\undefined}
}%
\providecommand \@ifnum [1]{%
 \ifnum #1\expandafter \@firstoftwo
 \else \expandafter \@secondoftwo
 \fi
}%
\providecommand \@ifx [1]{%
 \ifx #1\expandafter \@firstoftwo
 \else \expandafter \@secondoftwo
 \fi
}%
\providecommand \natexlab [1]{#1}%
\providecommand \enquote  [1]{``#1''}%
\providecommand \bibnamefont  [1]{#1}%
\providecommand \bibfnamefont [1]{#1}%
\providecommand \citenamefont [1]{#1}%
\providecommand \href@noop [0]{\@secondoftwo}%
\providecommand \href [0]{\begingroup \@sanitize@url \@href}%
\providecommand \@href[1]{\@@startlink{#1}\@@href}%
\providecommand \@@href[1]{\endgroup#1\@@endlink}%
\providecommand \@sanitize@url [0]{\catcode `\\12\catcode `\$12\catcode
  `\&12\catcode `\#12\catcode `\^12\catcode `\_12\catcode `\%12\relax}%
\providecommand \@@startlink[1]{}%
\providecommand \@@endlink[0]{}%
\providecommand \url  [0]{\begingroup\@sanitize@url \@url }%
\providecommand \@url [1]{\endgroup\@href {#1}{\urlprefix }}%
\providecommand \urlprefix  [0]{URL }%
\providecommand \Eprint [0]{\href }%
\providecommand \doibase [0]{http://dx.doi.org/}%
\providecommand \selectlanguage [0]{\@gobble}%
\providecommand \bibinfo  [0]{\@secondoftwo}%
\providecommand \bibfield  [0]{\@secondoftwo}%
\providecommand \translation [1]{[#1]}%
\providecommand \BibitemOpen [0]{}%
\providecommand \bibitemStop [0]{}%
\providecommand \bibitemNoStop [0]{.\EOS\space}%
\providecommand \EOS [0]{\spacefactor3000\relax}%
\providecommand \BibitemShut  [1]{\csname bibitem#1\endcsname}%
\let\auto@bib@innerbib\@empty
\bibitem [{\citenamefont {Roy}\ \emph {et~al.}(2017)\citenamefont {Roy},
  \citenamefont {Wilson},\ and\ \citenamefont {Firstenberg}}]{Roy-RMP2017}%
  \BibitemOpen
  \bibfield  {author} {\bibinfo {author} {\bibfnamefont {D.}~\bibnamefont
  {Roy}}, \bibinfo {author} {\bibfnamefont {C.~M.}\ \bibnamefont {Wilson}}, \
  and\ \bibinfo {author} {\bibfnamefont {O.}~\bibnamefont {Firstenberg}},\
  }\href {\doibase 10.1103/RevModPhys.89.021001} {\bibfield  {journal}
  {\bibinfo  {journal} {Rev. Mod. Phys.}\ }\textbf {\bibinfo {volume} {89}},\
  \bibinfo {pages} {021001} (\bibinfo {year} {2017})}\BibitemShut {NoStop}%
\bibitem [{\citenamefont {Gu}\ \emph {et~al.}(2017)\citenamefont {Gu},
  \citenamefont {Kockum}, \citenamefont {Miranowicz}, \citenamefont {Liu},\
  and\ \citenamefont {Nori}}]{Gu-PhysReports2017}%
  \BibitemOpen
  \bibfield  {author} {\bibinfo {author} {\bibfnamefont {X.}~\bibnamefont
  {Gu}}, \bibinfo {author} {\bibfnamefont {A.~F.}\ \bibnamefont {Kockum}},
  \bibinfo {author} {\bibfnamefont {A.}~\bibnamefont {Miranowicz}}, \bibinfo
  {author} {\bibfnamefont {Y.-x.}\ \bibnamefont {Liu}}, \ and\ \bibinfo
  {author} {\bibfnamefont {F.}~\bibnamefont {Nori}},\ }\href {\doibase
  https://doi.org/10.1016/j.physrep.2017.10.002} {\bibfield  {journal}
  {\bibinfo  {journal} {Phys. Rep.}\ }\textbf {\bibinfo {volume} {718-719}},\
  \bibinfo {pages} {1 } (\bibinfo {year} {2017})}\BibitemShut {NoStop}%
\bibitem [{\citenamefont {Sheremet}\ \emph {et~al.}(2023)\citenamefont
  {Sheremet}, \citenamefont {Petrov}, \citenamefont {Iorsh}, \citenamefont
  {Poshakinskiy},\ and\ \citenamefont {Poddubny}}]{Sheremet-RMP2023}%
  \BibitemOpen
  \bibfield  {author} {\bibinfo {author} {\bibfnamefont {A.~S.}\ \bibnamefont
  {Sheremet}}, \bibinfo {author} {\bibfnamefont {M.~I.}\ \bibnamefont
  {Petrov}}, \bibinfo {author} {\bibfnamefont {I.~V.}\ \bibnamefont {Iorsh}},
  \bibinfo {author} {\bibfnamefont {A.~V.}\ \bibnamefont {Poshakinskiy}}, \
  and\ \bibinfo {author} {\bibfnamefont {A.~N.}\ \bibnamefont {Poddubny}},\
  }\href {\doibase 10.1103/RevModPhys.95.015002} {\bibfield  {journal}
  {\bibinfo  {journal} {Rev. Mod. Phys.}\ }\textbf {\bibinfo {volume} {95}},\
  \bibinfo {pages} {015002} (\bibinfo {year} {2023})}\BibitemShut {NoStop}%
\bibitem [{\citenamefont {Shen}\ and\ \citenamefont
  {Fan}(2005{\natexlab{a}})}]{Shen-PRL2005}%
  \BibitemOpen
  \bibfield  {author} {\bibinfo {author} {\bibfnamefont {J.-T.}\ \bibnamefont
  {Shen}}\ and\ \bibinfo {author} {\bibfnamefont {S.}~\bibnamefont {Fan}},\
  }\href {\doibase 10.1103/PhysRevLett.95.213001} {\bibfield  {journal}
  {\bibinfo  {journal} {Phys. Rev. Lett.}\ }\textbf {\bibinfo {volume} {95}},\
  \bibinfo {pages} {213001} (\bibinfo {year} {2005}{\natexlab{a}})}\BibitemShut
  {NoStop}%
\bibitem [{\citenamefont {Chang}\ \emph {et~al.}(2006)\citenamefont {Chang},
  \citenamefont {S\o{}rensen}, \citenamefont {Hemmer},\ and\ \citenamefont
  {Lukin}}]{Chang-PRL2006}%
  \BibitemOpen
  \bibfield  {author} {\bibinfo {author} {\bibfnamefont {D.~E.}\ \bibnamefont
  {Chang}}, \bibinfo {author} {\bibfnamefont {A.~S.}\ \bibnamefont
  {S\o{}rensen}}, \bibinfo {author} {\bibfnamefont {P.~R.}\ \bibnamefont
  {Hemmer}}, \ and\ \bibinfo {author} {\bibfnamefont {M.~D.}\ \bibnamefont
  {Lukin}},\ }\href {\doibase 10.1103/PhysRevLett.97.053002} {\bibfield
  {journal} {\bibinfo  {journal} {Phys. Rev. Lett.}\ }\textbf {\bibinfo
  {volume} {97}},\ \bibinfo {pages} {053002} (\bibinfo {year}
  {2006})}\BibitemShut {NoStop}%
\bibitem [{\citenamefont {Shen}\ and\ \citenamefont
  {Fan}(2007)}]{Shen-PRL2007}%
  \BibitemOpen
  \bibfield  {author} {\bibinfo {author} {\bibfnamefont {J.-T.}\ \bibnamefont
  {Shen}}\ and\ \bibinfo {author} {\bibfnamefont {S.}~\bibnamefont {Fan}},\
  }\href {\doibase 10.1103/PhysRevLett.98.153003} {\bibfield  {journal}
  {\bibinfo  {journal} {Phys. Rev. Lett.}\ }\textbf {\bibinfo {volume} {98}},\
  \bibinfo {pages} {153003} (\bibinfo {year} {2007})}\BibitemShut {NoStop}%
\bibitem [{\citenamefont {Zhou}\ \emph {et~al.}(2008)\citenamefont {Zhou},
  \citenamefont {Gong}, \citenamefont {Liu}, \citenamefont {Sun},\ and\
  \citenamefont {Nori}}]{Zhou-PRL2008}%
  \BibitemOpen
  \bibfield  {author} {\bibinfo {author} {\bibfnamefont {L.}~\bibnamefont
  {Zhou}}, \bibinfo {author} {\bibfnamefont {Z.~R.}\ \bibnamefont {Gong}},
  \bibinfo {author} {\bibfnamefont {Y.-x.}\ \bibnamefont {Liu}}, \bibinfo
  {author} {\bibfnamefont {C.~P.}\ \bibnamefont {Sun}}, \ and\ \bibinfo
  {author} {\bibfnamefont {F.}~\bibnamefont {Nori}},\ }\href {\doibase
  10.1103/PhysRevLett.101.100501} {\bibfield  {journal} {\bibinfo  {journal}
  {Phys. Rev. Lett.}\ }\textbf {\bibinfo {volume} {101}},\ \bibinfo {pages}
  {100501} (\bibinfo {year} {2008})}\BibitemShut {NoStop}%
\bibitem [{\citenamefont {Astafiev}\ \emph {et~al.}(2010)\citenamefont
  {Astafiev}, \citenamefont {Zagoskin}, \citenamefont {Abdumalikov},
  \citenamefont {Pashkin}, \citenamefont {Yamamoto}, \citenamefont {Inomata},
  \citenamefont {Nakamura},\ and\ \citenamefont {Tsai}}]{Astafiev-Science2010}%
  \BibitemOpen
  \bibfield  {author} {\bibinfo {author} {\bibfnamefont {O.}~\bibnamefont
  {Astafiev}}, \bibinfo {author} {\bibfnamefont {A.~M.}\ \bibnamefont
  {Zagoskin}}, \bibinfo {author} {\bibfnamefont {A.~A.}\ \bibnamefont
  {Abdumalikov}}, \bibinfo {author} {\bibfnamefont {Y.~A.}\ \bibnamefont
  {Pashkin}}, \bibinfo {author} {\bibfnamefont {T.}~\bibnamefont {Yamamoto}},
  \bibinfo {author} {\bibfnamefont {K.}~\bibnamefont {Inomata}}, \bibinfo
  {author} {\bibfnamefont {Y.}~\bibnamefont {Nakamura}}, \ and\ \bibinfo
  {author} {\bibfnamefont {J.~S.}\ \bibnamefont {Tsai}},\ }\href {\doibase
  10.1126/science.1181918} {\bibfield  {journal} {\bibinfo  {journal}
  {Science}\ }\textbf {\bibinfo {volume} {327}},\ \bibinfo {pages} {840}
  (\bibinfo {year} {2010})}\BibitemShut {NoStop}%
\bibitem [{\citenamefont {Shi}\ \emph {et~al.}(2015)\citenamefont {Shi},
  \citenamefont {Chang},\ and\ \citenamefont {Cirac}}]{Shi-PRA2015}%
  \BibitemOpen
  \bibfield  {author} {\bibinfo {author} {\bibfnamefont {T.}~\bibnamefont
  {Shi}}, \bibinfo {author} {\bibfnamefont {D.~E.}\ \bibnamefont {Chang}}, \
  and\ \bibinfo {author} {\bibfnamefont {J.~I.}\ \bibnamefont {Cirac}},\ }\href
  {\doibase 10.1103/PhysRevA.92.053834} {\bibfield  {journal} {\bibinfo
  {journal} {Phys. Rev. A}\ }\textbf {\bibinfo {volume} {92}},\ \bibinfo
  {pages} {053834} (\bibinfo {year} {2015})}\BibitemShut {NoStop}%
\bibitem [{\citenamefont {Dicke}(1954)}]{PR-Dicke1954}%
  \BibitemOpen
  \bibfield  {author} {\bibinfo {author} {\bibfnamefont {R.~H.}\ \bibnamefont
  {Dicke}},\ }\href {\doibase 10.1103/PhysRev.93.99} {\bibfield  {journal}
  {\bibinfo  {journal} {Phys. Rev.}\ }\textbf {\bibinfo {volume} {93}},\
  \bibinfo {pages} {99} (\bibinfo {year} {1954})}\BibitemShut {NoStop}%
\bibitem [{\citenamefont {Vetter}\ \emph {et~al.}(2016)\citenamefont {Vetter},
  \citenamefont {Wang}, \citenamefont {Wang},\ and\ \citenamefont
  {Scully}}]{Vetter-PScr2016}%
  \BibitemOpen
  \bibfield  {author} {\bibinfo {author} {\bibfnamefont {P.~A.}\ \bibnamefont
  {Vetter}}, \bibinfo {author} {\bibfnamefont {L.}~\bibnamefont {Wang}},
  \bibinfo {author} {\bibfnamefont {D.-W.}\ \bibnamefont {Wang}}, \ and\
  \bibinfo {author} {\bibfnamefont {M.~O.}\ \bibnamefont {Scully}},\ }\href
  {\doibase 10.1088/0031-8949/91/2/023007} {\bibfield  {journal} {\bibinfo
  {journal} {Physica Scripta}\ }\textbf {\bibinfo {volume} {91}},\ \bibinfo
  {pages} {023007} (\bibinfo {year} {2016})}\BibitemShut {NoStop}%
\bibitem [{\citenamefont {van Loo}\ \emph {et~al.}(2013)\citenamefont {van
  Loo}, \citenamefont {Fedorov}, \citenamefont {Lalumière}, \citenamefont
  {Sanders}, \citenamefont {Blais},\ and\ \citenamefont
  {Wallraff}}]{Loo-Science2013}%
  \BibitemOpen
  \bibfield  {author} {\bibinfo {author} {\bibfnamefont {A.~F.}\ \bibnamefont
  {van Loo}}, \bibinfo {author} {\bibfnamefont {A.}~\bibnamefont {Fedorov}},
  \bibinfo {author} {\bibfnamefont {K.}~\bibnamefont {Lalumière}}, \bibinfo
  {author} {\bibfnamefont {B.~C.}\ \bibnamefont {Sanders}}, \bibinfo {author}
  {\bibfnamefont {A.}~\bibnamefont {Blais}}, \ and\ \bibinfo {author}
  {\bibfnamefont {A.}~\bibnamefont {Wallraff}},\ }\href {\doibase
  10.1126/science.1244324} {\bibfield  {journal} {\bibinfo  {journal}
  {Science}\ }\textbf {\bibinfo {volume} {342}},\ \bibinfo {pages} {1494}
  (\bibinfo {year} {2013})}\BibitemShut {NoStop}%
\bibitem [{\citenamefont {Zhang}\ and\ \citenamefont
  {M\o{}lmer}(2019)}]{Zhang-PRL2019}%
  \BibitemOpen
  \bibfield  {author} {\bibinfo {author} {\bibfnamefont {Y.-X.}\ \bibnamefont
  {Zhang}}\ and\ \bibinfo {author} {\bibfnamefont {K.}~\bibnamefont
  {M\o{}lmer}},\ }\href {\doibase 10.1103/PhysRevLett.122.203605} {\bibfield
  {journal} {\bibinfo  {journal} {Phys. Rev. Lett.}\ }\textbf {\bibinfo
  {volume} {122}},\ \bibinfo {pages} {203605} (\bibinfo {year}
  {2019})}\BibitemShut {NoStop}%
\bibitem [{\citenamefont {Ke}\ \emph {et~al.}(2019)\citenamefont {Ke},
  \citenamefont {Poshakinskiy}, \citenamefont {Lee}, \citenamefont {Kivshar},\
  and\ \citenamefont {Poddubny}}]{Ke-PRL2019}%
  \BibitemOpen
  \bibfield  {author} {\bibinfo {author} {\bibfnamefont {Y.}~\bibnamefont
  {Ke}}, \bibinfo {author} {\bibfnamefont {A.~V.}\ \bibnamefont
  {Poshakinskiy}}, \bibinfo {author} {\bibfnamefont {C.}~\bibnamefont {Lee}},
  \bibinfo {author} {\bibfnamefont {Y.~S.}\ \bibnamefont {Kivshar}}, \ and\
  \bibinfo {author} {\bibfnamefont {A.~N.}\ \bibnamefont {Poddubny}},\ }\href
  {\doibase 10.1103/PhysRevLett.123.253601} {\bibfield  {journal} {\bibinfo
  {journal} {Phys. Rev. Lett.}\ }\textbf {\bibinfo {volume} {123}},\ \bibinfo
  {pages} {253601} (\bibinfo {year} {2019})}\BibitemShut {NoStop}%
\bibitem [{\citenamefont {Wang}\ \emph {et~al.}(2020)\citenamefont {Wang},
  \citenamefont {Li}, \citenamefont {Feng}, \citenamefont {Song}, \citenamefont
  {Song}, \citenamefont {Liu}, \citenamefont {Guo}, \citenamefont {Zhang},
  \citenamefont {Dong}, \citenamefont {Zheng}, \citenamefont {Wang},\ and\
  \citenamefont {Wang}}]{Wang-PRL2020}%
  \BibitemOpen
  \bibfield  {author} {\bibinfo {author} {\bibfnamefont {Z.}~\bibnamefont
  {Wang}}, \bibinfo {author} {\bibfnamefont {H.}~\bibnamefont {Li}}, \bibinfo
  {author} {\bibfnamefont {W.}~\bibnamefont {Feng}}, \bibinfo {author}
  {\bibfnamefont {X.}~\bibnamefont {Song}}, \bibinfo {author} {\bibfnamefont
  {C.}~\bibnamefont {Song}}, \bibinfo {author} {\bibfnamefont {W.}~\bibnamefont
  {Liu}}, \bibinfo {author} {\bibfnamefont {Q.}~\bibnamefont {Guo}}, \bibinfo
  {author} {\bibfnamefont {X.}~\bibnamefont {Zhang}}, \bibinfo {author}
  {\bibfnamefont {H.}~\bibnamefont {Dong}}, \bibinfo {author} {\bibfnamefont
  {D.}~\bibnamefont {Zheng}}, \bibinfo {author} {\bibfnamefont
  {H.}~\bibnamefont {Wang}}, \ and\ \bibinfo {author} {\bibfnamefont {D.-W.}\
  \bibnamefont {Wang}},\ }\href {\doibase 10.1103/PhysRevLett.124.013601}
  {\bibfield  {journal} {\bibinfo  {journal} {Phys. Rev. Lett.}\ }\textbf
  {\bibinfo {volume} {124}},\ \bibinfo {pages} {013601} (\bibinfo {year}
  {2020})}\BibitemShut {NoStop}%
\bibitem [{\citenamefont {Zheng}\ and\ \citenamefont
  {Baranger}(2013)}]{Zheng-PRL2013}%
  \BibitemOpen
  \bibfield  {author} {\bibinfo {author} {\bibfnamefont {H.}~\bibnamefont
  {Zheng}}\ and\ \bibinfo {author} {\bibfnamefont {H.~U.}\ \bibnamefont
  {Baranger}},\ }\href {\doibase 10.1103/PhysRevLett.110.113601} {\bibfield
  {journal} {\bibinfo  {journal} {Phys. Rev. Lett.}\ }\textbf {\bibinfo
  {volume} {110}},\ \bibinfo {pages} {113601} (\bibinfo {year}
  {2013})}\BibitemShut {NoStop}%
\bibitem [{\citenamefont {Gonzalez-Ballestero}\ \emph
  {et~al.}(2014)\citenamefont {Gonzalez-Ballestero}, \citenamefont {Moreno},\
  and\ \citenamefont {Garcia-Vidal}}]{Ballestero-PRA2014}%
  \BibitemOpen
  \bibfield  {author} {\bibinfo {author} {\bibfnamefont {C.}~\bibnamefont
  {Gonzalez-Ballestero}}, \bibinfo {author} {\bibfnamefont {E.}~\bibnamefont
  {Moreno}}, \ and\ \bibinfo {author} {\bibfnamefont {F.~J.}\ \bibnamefont
  {Garcia-Vidal}},\ }\href {\doibase 10.1103/PhysRevA.89.042328} {\bibfield
  {journal} {\bibinfo  {journal} {Phys. Rev. A}\ }\textbf {\bibinfo {volume}
  {89}},\ \bibinfo {pages} {042328} (\bibinfo {year} {2014})}\BibitemShut
  {NoStop}%
\bibitem [{\citenamefont {Facchi}\ \emph {et~al.}(2016)\citenamefont {Facchi},
  \citenamefont {Kim}, \citenamefont {Pascazio}, \citenamefont {Pepe},
  \citenamefont {Pomarico},\ and\ \citenamefont {Tufarelli}}]{Facchi-PRA-2016}%
  \BibitemOpen
  \bibfield  {author} {\bibinfo {author} {\bibfnamefont {P.}~\bibnamefont
  {Facchi}}, \bibinfo {author} {\bibfnamefont {M.~S.}\ \bibnamefont {Kim}},
  \bibinfo {author} {\bibfnamefont {S.}~\bibnamefont {Pascazio}}, \bibinfo
  {author} {\bibfnamefont {F.~V.}\ \bibnamefont {Pepe}}, \bibinfo {author}
  {\bibfnamefont {D.}~\bibnamefont {Pomarico}}, \ and\ \bibinfo {author}
  {\bibfnamefont {T.}~\bibnamefont {Tufarelli}},\ }\href {\doibase
  10.1103/PhysRevA.94.043839} {\bibfield  {journal} {\bibinfo  {journal} {Phys.
  Rev. A}\ }\textbf {\bibinfo {volume} {94}},\ \bibinfo {pages} {043839}
  (\bibinfo {year} {2016})}\BibitemShut {NoStop}%
\bibitem [{\citenamefont {Mirza}\ and\ \citenamefont
  {Schotland}(2016)}]{Mirza-PRA2016}%
  \BibitemOpen
  \bibfield  {author} {\bibinfo {author} {\bibfnamefont {I.~M.}\ \bibnamefont
  {Mirza}}\ and\ \bibinfo {author} {\bibfnamefont {J.~C.}\ \bibnamefont
  {Schotland}},\ }\href {\doibase 10.1103/PhysRevA.94.012302} {\bibfield
  {journal} {\bibinfo  {journal} {Phys. Rev. A}\ }\textbf {\bibinfo {volume}
  {94}},\ \bibinfo {pages} {012302} (\bibinfo {year} {2016})}\BibitemShut
  {NoStop}%
\bibitem [{\citenamefont {Chang}\ \emph {et~al.}(2012)\citenamefont {Chang},
  \citenamefont {Jiang}, \citenamefont {Gorshkov},\ and\ \citenamefont
  {Kimble}}]{Chang-NJP2012}%
  \BibitemOpen
  \bibfield  {author} {\bibinfo {author} {\bibfnamefont {D.~E.}\ \bibnamefont
  {Chang}}, \bibinfo {author} {\bibfnamefont {L.}~\bibnamefont {Jiang}},
  \bibinfo {author} {\bibfnamefont {A.~V.}\ \bibnamefont {Gorshkov}}, \ and\
  \bibinfo {author} {\bibfnamefont {H.~J.}\ \bibnamefont {Kimble}},\ }\href
  {\doibase 10.1088/1367-2630/14/6/063003} {\bibfield  {journal} {\bibinfo
  {journal} {New J. Phys.}\ }\textbf {\bibinfo {volume} {14}},\ \bibinfo
  {pages} {063003} (\bibinfo {year} {2012})}\BibitemShut {NoStop}%
\bibitem [{\citenamefont {Mirhosseini}\ \emph {et~al.}(2019)\citenamefont
  {Mirhosseini}, \citenamefont {Kim}, \citenamefont {Zhang}, \citenamefont
  {Sipahigil}, \citenamefont {Dieterle}, \citenamefont {Keller}, \citenamefont
  {Asenjo-Garcia}, \citenamefont {Chang},\ and\ \citenamefont
  {Painter}}]{Mirhosseini-Natrue2019}%
  \BibitemOpen
  \bibfield  {author} {\bibinfo {author} {\bibfnamefont {M.}~\bibnamefont
  {Mirhosseini}}, \bibinfo {author} {\bibfnamefont {E.}~\bibnamefont {Kim}},
  \bibinfo {author} {\bibfnamefont {X.}~\bibnamefont {Zhang}}, \bibinfo
  {author} {\bibfnamefont {A.}~\bibnamefont {Sipahigil}}, \bibinfo {author}
  {\bibfnamefont {P.~B.}\ \bibnamefont {Dieterle}}, \bibinfo {author}
  {\bibfnamefont {A.~J.}\ \bibnamefont {Keller}}, \bibinfo {author}
  {\bibfnamefont {A.}~\bibnamefont {Asenjo-Garcia}}, \bibinfo {author}
  {\bibfnamefont {D.~E.}\ \bibnamefont {Chang}}, \ and\ \bibinfo {author}
  {\bibfnamefont {O.}~\bibnamefont {Painter}},\ }\href {\doibase
  10.1038/s41586-019-1196-1} {\bibfield  {journal} {\bibinfo  {journal} {Nature
  (London)}\ }\textbf {\bibinfo {volume} {569}},\ \bibinfo {pages} {692}
  (\bibinfo {year} {2019})}\BibitemShut {NoStop}%
\bibitem [{\citenamefont {Nie}\ \emph {et~al.}(2023)\citenamefont {Nie},
  \citenamefont {Shi}, \citenamefont {Liu},\ and\ \citenamefont
  {Nori}}]{Nie-PRL2023}%
  \BibitemOpen
  \bibfield  {author} {\bibinfo {author} {\bibfnamefont {W.}~\bibnamefont
  {Nie}}, \bibinfo {author} {\bibfnamefont {T.}~\bibnamefont {Shi}}, \bibinfo
  {author} {\bibfnamefont {Y.-x.}\ \bibnamefont {Liu}}, \ and\ \bibinfo
  {author} {\bibfnamefont {F.}~\bibnamefont {Nori}},\ }\href {\doibase
  10.1103/PhysRevLett.131.103602} {\bibfield  {journal} {\bibinfo  {journal}
  {Phys. Rev. Lett.}\ }\textbf {\bibinfo {volume} {131}},\ \bibinfo {pages}
  {103602} (\bibinfo {year} {2023})}\BibitemShut {NoStop}%
\bibitem [{\citenamefont {Kockum}\ \emph {et~al.}(2018)\citenamefont {Kockum},
  \citenamefont {Johansson},\ and\ \citenamefont {Nori}}]{Kockum-PRL2018}%
  \BibitemOpen
  \bibfield  {author} {\bibinfo {author} {\bibfnamefont {A.~F.}\ \bibnamefont
  {Kockum}}, \bibinfo {author} {\bibfnamefont {G.}~\bibnamefont {Johansson}}, \
  and\ \bibinfo {author} {\bibfnamefont {F.}~\bibnamefont {Nori}},\ }\href
  {\doibase 10.1103/PhysRevLett.120.140404} {\bibfield  {journal} {\bibinfo
  {journal} {Phys. Rev. Lett.}\ }\textbf {\bibinfo {volume} {120}},\ \bibinfo
  {pages} {140404} (\bibinfo {year} {2018})}\BibitemShut {NoStop}%
\bibitem [{\citenamefont {Nie}\ \emph {et~al.}(2021)\citenamefont {Nie},
  \citenamefont {Shi}, \citenamefont {Nori},\ and\ \citenamefont
  {Liu}}]{Nie-PRApplied2021}%
  \BibitemOpen
  \bibfield  {author} {\bibinfo {author} {\bibfnamefont {W.}~\bibnamefont
  {Nie}}, \bibinfo {author} {\bibfnamefont {T.}~\bibnamefont {Shi}}, \bibinfo
  {author} {\bibfnamefont {F.}~\bibnamefont {Nori}}, \ and\ \bibinfo {author}
  {\bibfnamefont {Y.-x.}\ \bibnamefont {Liu}},\ }\href {\doibase
  10.1103/PhysRevApplied.15.044041} {\bibfield  {journal} {\bibinfo  {journal}
  {Phys. Rev. Appl.}\ }\textbf {\bibinfo {volume} {15}},\ \bibinfo {pages}
  {044041} (\bibinfo {year} {2021})}\BibitemShut {NoStop}%
\bibitem [{\citenamefont {Tsoi}\ and\ \citenamefont
  {Law}(2008)}]{Tsoi-PRA2008}%
  \BibitemOpen
  \bibfield  {author} {\bibinfo {author} {\bibfnamefont {T.~S.}\ \bibnamefont
  {Tsoi}}\ and\ \bibinfo {author} {\bibfnamefont {C.~K.}\ \bibnamefont {Law}},\
  }\href {\doibase 10.1103/PhysRevA.78.063832} {\bibfield  {journal} {\bibinfo
  {journal} {Phys. Rev. A}\ }\textbf {\bibinfo {volume} {78}},\ \bibinfo
  {pages} {063832} (\bibinfo {year} {2008})}\BibitemShut {NoStop}%
\bibitem [{\citenamefont {Liao}\ \emph {et~al.}(2015)\citenamefont {Liao},
  \citenamefont {Zeng}, \citenamefont {Zhu},\ and\ \citenamefont
  {Zubairy}}]{Liao-PRA2015}%
  \BibitemOpen
  \bibfield  {author} {\bibinfo {author} {\bibfnamefont {Z.}~\bibnamefont
  {Liao}}, \bibinfo {author} {\bibfnamefont {X.}~\bibnamefont {Zeng}}, \bibinfo
  {author} {\bibfnamefont {S.-Y.}\ \bibnamefont {Zhu}}, \ and\ \bibinfo
  {author} {\bibfnamefont {M.~S.}\ \bibnamefont {Zubairy}},\ }\href {\doibase
  10.1103/PhysRevA.92.023806} {\bibfield  {journal} {\bibinfo  {journal} {Phys.
  Rev. A}\ }\textbf {\bibinfo {volume} {92}},\ \bibinfo {pages} {023806}
  (\bibinfo {year} {2015})}\BibitemShut {NoStop}%
\bibitem [{\citenamefont {Cheng}\ \emph {et~al.}(2017)\citenamefont {Cheng},
  \citenamefont {Xu},\ and\ \citenamefont {Agarwal}}]{Cheng-PRA2017}%
  \BibitemOpen
  \bibfield  {author} {\bibinfo {author} {\bibfnamefont {M.-T.}\ \bibnamefont
  {Cheng}}, \bibinfo {author} {\bibfnamefont {J.}~\bibnamefont {Xu}}, \ and\
  \bibinfo {author} {\bibfnamefont {G.~S.}\ \bibnamefont {Agarwal}},\ }\href
  {\doibase 10.1103/PhysRevA.95.053807} {\bibfield  {journal} {\bibinfo
  {journal} {Phys. Rev. A}\ }\textbf {\bibinfo {volume} {95}},\ \bibinfo
  {pages} {053807} (\bibinfo {year} {2017})}\BibitemShut {NoStop}%
\bibitem [{\citenamefont {Ruostekoski}\ and\ \citenamefont
  {Javanainen}(2017)}]{Ruostekoski-PRA2017}%
  \BibitemOpen
  \bibfield  {author} {\bibinfo {author} {\bibfnamefont {J.}~\bibnamefont
  {Ruostekoski}}\ and\ \bibinfo {author} {\bibfnamefont {J.}~\bibnamefont
  {Javanainen}},\ }\href {\doibase 10.1103/PhysRevA.96.033857} {\bibfield
  {journal} {\bibinfo  {journal} {Phys. Rev. A}\ }\textbf {\bibinfo {volume}
  {96}},\ \bibinfo {pages} {033857} (\bibinfo {year} {2017})}\BibitemShut
  {NoStop}%
\bibitem [{\citenamefont {Mukhopadhyay}\ and\ \citenamefont
  {Agarwal}(2019)}]{Mukhopadhyay-PRA2019}%
  \BibitemOpen
  \bibfield  {author} {\bibinfo {author} {\bibfnamefont {D.}~\bibnamefont
  {Mukhopadhyay}}\ and\ \bibinfo {author} {\bibfnamefont {G.~S.}\ \bibnamefont
  {Agarwal}},\ }\href {\doibase 10.1103/PhysRevA.100.013812} {\bibfield
  {journal} {\bibinfo  {journal} {Phys. Rev. A}\ }\textbf {\bibinfo {volume}
  {100}},\ \bibinfo {pages} {013812} (\bibinfo {year} {2019})}\BibitemShut
  {NoStop}%
\bibitem [{\citenamefont {Mukhopadhyay}\ and\ \citenamefont
  {Agarwal}(2020)}]{Mukhopadhyay-PRA2020}%
  \BibitemOpen
  \bibfield  {author} {\bibinfo {author} {\bibfnamefont {D.}~\bibnamefont
  {Mukhopadhyay}}\ and\ \bibinfo {author} {\bibfnamefont {G.~S.}\ \bibnamefont
  {Agarwal}},\ }\href {\doibase 10.1103/PhysRevA.101.063814} {\bibfield
  {journal} {\bibinfo  {journal} {Phys. Rev. A}\ }\textbf {\bibinfo {volume}
  {101}},\ \bibinfo {pages} {063814} (\bibinfo {year} {2020})}\BibitemShut
  {NoStop}%
\bibitem [{\citenamefont {Ask}\ \emph {et~al.}()\citenamefont {Ask},
  \citenamefont {Fang},\ and\ \citenamefont {Kockum}}]{Ask-arXive2020}%
  \BibitemOpen
  \bibfield  {author} {\bibinfo {author} {\bibfnamefont {A.}~\bibnamefont
  {Ask}}, \bibinfo {author} {\bibfnamefont {Y.-L.~L.}\ \bibnamefont {Fang}}, \
  and\ \bibinfo {author} {\bibfnamefont {A.~F.}\ \bibnamefont {Kockum}},\
  }\href {https://arxiv.org/abs/2011.15077} {}\Eprint
  {http://arxiv.org/abs/2011.15077} {arXiv:2011.15077} \BibitemShut {NoStop}%
\bibitem [{\citenamefont {Jia}\ and\ \citenamefont {Cai}(2022)}]{Jia-EPJP2022}%
  \BibitemOpen
  \bibfield  {author} {\bibinfo {author} {\bibfnamefont {W.~Z.}\ \bibnamefont
  {Jia}}\ and\ \bibinfo {author} {\bibfnamefont {Q.~Y.}\ \bibnamefont {Cai}},\
  }\href {\doibase 10.1140/epjp/s13360-022-03284-4} {\bibfield  {journal}
  {\bibinfo  {journal} {Eur. Phys. J. Plus}\ }\textbf {\bibinfo {volume}
  {137}},\ \bibinfo {pages} {1082} (\bibinfo {year} {2022})}\BibitemShut
  {NoStop}%
\bibitem [{\citenamefont {Feng}\ and\ \citenamefont
  {Jia}(2021)}]{Feng-PRA2021}%
  \BibitemOpen
  \bibfield  {author} {\bibinfo {author} {\bibfnamefont {S.~L.}\ \bibnamefont
  {Feng}}\ and\ \bibinfo {author} {\bibfnamefont {W.~Z.}\ \bibnamefont {Jia}},\
  }\href {\doibase 10.1103/PhysRevA.104.063712} {\bibfield  {journal} {\bibinfo
   {journal} {Phys. Rev. A}\ }\textbf {\bibinfo {volume} {104}},\ \bibinfo
  {pages} {063712} (\bibinfo {year} {2021})}\BibitemShut {NoStop}%
\bibitem [{\citenamefont {Bendickson}\ \emph {et~al.}(1996)\citenamefont
  {Bendickson}, \citenamefont {Dowling},\ and\ \citenamefont
  {Scalora}}]{Bendickson-PRE1996}%
  \BibitemOpen
  \bibfield  {author} {\bibinfo {author} {\bibfnamefont {J.~M.}\ \bibnamefont
  {Bendickson}}, \bibinfo {author} {\bibfnamefont {J.~P.}\ \bibnamefont
  {Dowling}}, \ and\ \bibinfo {author} {\bibfnamefont {M.}~\bibnamefont
  {Scalora}},\ }\href {\doibase 10.1103/PhysRevE.53.4107} {\bibfield  {journal}
  {\bibinfo  {journal} {Phys. Rev. E}\ }\textbf {\bibinfo {volume} {53}},\
  \bibinfo {pages} {4107} (\bibinfo {year} {1996})}\BibitemShut {NoStop}%
\bibitem [{\citenamefont {Hood}\ \emph {et~al.}(2016)\citenamefont {Hood},
  \citenamefont {Goban}, \citenamefont {Asenjo-Garcia}, \citenamefont {Lu},
  \citenamefont {Yu}, \citenamefont {Chang},\ and\ \citenamefont
  {Kimble}}]{Hood-PNAS2016}%
  \BibitemOpen
  \bibfield  {author} {\bibinfo {author} {\bibfnamefont {J.~D.}\ \bibnamefont
  {Hood}}, \bibinfo {author} {\bibfnamefont {A.}~\bibnamefont {Goban}},
  \bibinfo {author} {\bibfnamefont {A.}~\bibnamefont {Asenjo-Garcia}}, \bibinfo
  {author} {\bibfnamefont {M.}~\bibnamefont {Lu}}, \bibinfo {author}
  {\bibfnamefont {S.-P.}\ \bibnamefont {Yu}}, \bibinfo {author} {\bibfnamefont
  {D.~E.}\ \bibnamefont {Chang}}, \ and\ \bibinfo {author} {\bibfnamefont
  {H.~J.}\ \bibnamefont {Kimble}},\ }\href {\doibase 10.1073/pnas.1603788113}
  {\bibfield  {journal} {\bibinfo  {journal} {Proc. Natl. Acad. Sci.}\ }\textbf
  {\bibinfo {volume} {113}},\ \bibinfo {pages} {10507} (\bibinfo {year}
  {2016})}\BibitemShut {NoStop}%
\bibitem [{\citenamefont {Liu}\ and\ \citenamefont
  {Houck}(2017)}]{Liu-NatPhys2017}%
  \BibitemOpen
  \bibfield  {author} {\bibinfo {author} {\bibfnamefont {Y.}~\bibnamefont
  {Liu}}\ and\ \bibinfo {author} {\bibfnamefont {A.~A.}\ \bibnamefont
  {Houck}},\ }\href {\doibase 10.1038/nphys3834} {\bibfield  {journal}
  {\bibinfo  {journal} {Nat. Phys.}\ }\textbf {\bibinfo {volume} {13}},\
  \bibinfo {pages} {48} (\bibinfo {year} {2017})}\BibitemShut {NoStop}%
\bibitem [{\citenamefont {Sundaresan}\ \emph {et~al.}(2019)\citenamefont
  {Sundaresan}, \citenamefont {Lundgren}, \citenamefont {Zhu}, \citenamefont
  {Gorshkov},\ and\ \citenamefont {Houck}}]{Sundaresan-PRX2019}%
  \BibitemOpen
  \bibfield  {author} {\bibinfo {author} {\bibfnamefont {N.~M.}\ \bibnamefont
  {Sundaresan}}, \bibinfo {author} {\bibfnamefont {R.}~\bibnamefont
  {Lundgren}}, \bibinfo {author} {\bibfnamefont {G.}~\bibnamefont {Zhu}},
  \bibinfo {author} {\bibfnamefont {A.~V.}\ \bibnamefont {Gorshkov}}, \ and\
  \bibinfo {author} {\bibfnamefont {A.~A.}\ \bibnamefont {Houck}},\ }\href
  {\doibase 10.1103/PhysRevX.9.011021} {\bibfield  {journal} {\bibinfo
  {journal} {Phys. Rev. X}\ }\textbf {\bibinfo {volume} {9}},\ \bibinfo {pages}
  {011021} (\bibinfo {year} {2019})}\BibitemShut {NoStop}%
\bibitem [{\citenamefont {Kim}\ \emph {et~al.}(2021)\citenamefont {Kim},
  \citenamefont {Zhang}, \citenamefont {Ferreira}, \citenamefont {Banker},
  \citenamefont {Iverson}, \citenamefont {Sipahigil}, \citenamefont {Bello},
  \citenamefont {Gonz\'alez-Tudela}, \citenamefont {Mirhosseini},\ and\
  \citenamefont {Painter}}]{Kim-PRX2021}%
  \BibitemOpen
  \bibfield  {author} {\bibinfo {author} {\bibfnamefont {E.}~\bibnamefont
  {Kim}}, \bibinfo {author} {\bibfnamefont {X.}~\bibnamefont {Zhang}}, \bibinfo
  {author} {\bibfnamefont {V.~S.}\ \bibnamefont {Ferreira}}, \bibinfo {author}
  {\bibfnamefont {J.}~\bibnamefont {Banker}}, \bibinfo {author} {\bibfnamefont
  {J.~K.}\ \bibnamefont {Iverson}}, \bibinfo {author} {\bibfnamefont
  {A.}~\bibnamefont {Sipahigil}}, \bibinfo {author} {\bibfnamefont
  {M.}~\bibnamefont {Bello}}, \bibinfo {author} {\bibfnamefont
  {A.}~\bibnamefont {Gonz\'alez-Tudela}}, \bibinfo {author} {\bibfnamefont
  {M.}~\bibnamefont {Mirhosseini}}, \ and\ \bibinfo {author} {\bibfnamefont
  {O.}~\bibnamefont {Painter}},\ }\href {\doibase 10.1103/PhysRevX.11.011015}
  {\bibfield  {journal} {\bibinfo  {journal} {Phys. Rev. X}\ }\textbf {\bibinfo
  {volume} {11}},\ \bibinfo {pages} {011015} (\bibinfo {year}
  {2021})}\BibitemShut {NoStop}%
\bibitem [{\citenamefont {Ferreira}\ \emph {et~al.}(2021)\citenamefont
  {Ferreira}, \citenamefont {Banker}, \citenamefont {Sipahigil}, \citenamefont
  {Matheny}, \citenamefont {Keller}, \citenamefont {Kim}, \citenamefont
  {Mirhosseini},\ and\ \citenamefont {Painter}}]{Ferreira-PRX2021}%
  \BibitemOpen
  \bibfield  {author} {\bibinfo {author} {\bibfnamefont {V.~S.}\ \bibnamefont
  {Ferreira}}, \bibinfo {author} {\bibfnamefont {J.}~\bibnamefont {Banker}},
  \bibinfo {author} {\bibfnamefont {A.}~\bibnamefont {Sipahigil}}, \bibinfo
  {author} {\bibfnamefont {M.~H.}\ \bibnamefont {Matheny}}, \bibinfo {author}
  {\bibfnamefont {A.~J.}\ \bibnamefont {Keller}}, \bibinfo {author}
  {\bibfnamefont {E.}~\bibnamefont {Kim}}, \bibinfo {author} {\bibfnamefont
  {M.}~\bibnamefont {Mirhosseini}}, \ and\ \bibinfo {author} {\bibfnamefont
  {O.}~\bibnamefont {Painter}},\ }\href {\doibase 10.1103/PhysRevX.11.041043}
  {\bibfield  {journal} {\bibinfo  {journal} {Phys. Rev. X}\ }\textbf {\bibinfo
  {volume} {11}},\ \bibinfo {pages} {041043} (\bibinfo {year}
  {2021})}\BibitemShut {NoStop}%
\bibitem [{\citenamefont {Scigliuzzo}\ \emph {et~al.}(2022)\citenamefont
  {Scigliuzzo}, \citenamefont {Calaj\`o}, \citenamefont {Ciccarello},
  \citenamefont {Perez~Lozano}, \citenamefont {Bengtsson}, \citenamefont
  {Scarlino}, \citenamefont {Wallraff}, \citenamefont {Chang}, \citenamefont
  {Delsing},\ and\ \citenamefont {Gasparinetti}}]{Scigliuzzo-PRX2021}%
  \BibitemOpen
  \bibfield  {author} {\bibinfo {author} {\bibfnamefont {M.}~\bibnamefont
  {Scigliuzzo}}, \bibinfo {author} {\bibfnamefont {G.}~\bibnamefont
  {Calaj\`o}}, \bibinfo {author} {\bibfnamefont {F.}~\bibnamefont
  {Ciccarello}}, \bibinfo {author} {\bibfnamefont {D.}~\bibnamefont
  {Perez~Lozano}}, \bibinfo {author} {\bibfnamefont {A.}~\bibnamefont
  {Bengtsson}}, \bibinfo {author} {\bibfnamefont {P.}~\bibnamefont {Scarlino}},
  \bibinfo {author} {\bibfnamefont {A.}~\bibnamefont {Wallraff}}, \bibinfo
  {author} {\bibfnamefont {D.}~\bibnamefont {Chang}}, \bibinfo {author}
  {\bibfnamefont {P.}~\bibnamefont {Delsing}}, \ and\ \bibinfo {author}
  {\bibfnamefont {S.}~\bibnamefont {Gasparinetti}},\ }\href {\doibase
  10.1103/PhysRevX.12.031036} {\bibfield  {journal} {\bibinfo  {journal} {Phys.
  Rev. X}\ }\textbf {\bibinfo {volume} {12}},\ \bibinfo {pages} {031036}
  (\bibinfo {year} {2022})}\BibitemShut {NoStop}%
\bibitem [{\citenamefont {Zhang}\ \emph {et~al.}(2023)\citenamefont {Zhang},
  \citenamefont {Kim}, \citenamefont {Mark}, \citenamefont {Choi},\ and\
  \citenamefont {Painter}}]{Zhang-Science2023}%
  \BibitemOpen
  \bibfield  {author} {\bibinfo {author} {\bibfnamefont {X.}~\bibnamefont
  {Zhang}}, \bibinfo {author} {\bibfnamefont {E.}~\bibnamefont {Kim}}, \bibinfo
  {author} {\bibfnamefont {D.~K.}\ \bibnamefont {Mark}}, \bibinfo {author}
  {\bibfnamefont {S.}~\bibnamefont {Choi}}, \ and\ \bibinfo {author}
  {\bibfnamefont {O.}~\bibnamefont {Painter}},\ }\href {\doibase
  10.1126/science.ade7651} {\bibfield  {journal} {\bibinfo  {journal}
  {Science}\ }\textbf {\bibinfo {volume} {379}},\ \bibinfo {pages} {278}
  (\bibinfo {year} {2023})}\BibitemShut {NoStop}%
\bibitem [{\citenamefont {Rakhmanov}\ \emph {et~al.}(2008)\citenamefont
  {Rakhmanov}, \citenamefont {Zagoskin}, \citenamefont {Savel'ev},\ and\
  \citenamefont {Nori}}]{Rakhmanov-PRB2008}%
  \BibitemOpen
  \bibfield  {author} {\bibinfo {author} {\bibfnamefont {A.~L.}\ \bibnamefont
  {Rakhmanov}}, \bibinfo {author} {\bibfnamefont {A.~M.}\ \bibnamefont
  {Zagoskin}}, \bibinfo {author} {\bibfnamefont {S.}~\bibnamefont {Savel'ev}},
  \ and\ \bibinfo {author} {\bibfnamefont {F.}~\bibnamefont {Nori}},\ }\href
  {\doibase 10.1103/PhysRevB.77.144507} {\bibfield  {journal} {\bibinfo
  {journal} {Phys. Rev. B}\ }\textbf {\bibinfo {volume} {77}},\ \bibinfo
  {pages} {144507} (\bibinfo {year} {2008})}\BibitemShut {NoStop}%
\bibitem [{\citenamefont {Hutter}\ \emph {et~al.}(2011)\citenamefont {Hutter},
  \citenamefont {Thol\'en}, \citenamefont {Stannigel}, \citenamefont {Lidmar},\
  and\ \citenamefont {Haviland}}]{Hutter-PRB2011}%
  \BibitemOpen
  \bibfield  {author} {\bibinfo {author} {\bibfnamefont {C.}~\bibnamefont
  {Hutter}}, \bibinfo {author} {\bibfnamefont {E.~A.}\ \bibnamefont
  {Thol\'en}}, \bibinfo {author} {\bibfnamefont {K.}~\bibnamefont {Stannigel}},
  \bibinfo {author} {\bibfnamefont {J.}~\bibnamefont {Lidmar}}, \ and\ \bibinfo
  {author} {\bibfnamefont {D.~B.}\ \bibnamefont {Haviland}},\ }\href {\doibase
  10.1103/PhysRevB.83.014511} {\bibfield  {journal} {\bibinfo  {journal} {Phys.
  Rev. B}\ }\textbf {\bibinfo {volume} {83}},\ \bibinfo {pages} {014511}
  (\bibinfo {year} {2011})}\BibitemShut {NoStop}%
\bibitem [{\citenamefont {Zueco}\ \emph {et~al.}(2012)\citenamefont {Zueco},
  \citenamefont {Mazo}, \citenamefont {Solano},\ and\ \citenamefont
  {Garc\'{\i}a-Ripoll}}]{Zueco-2PRB012}%
  \BibitemOpen
  \bibfield  {author} {\bibinfo {author} {\bibfnamefont {D.}~\bibnamefont
  {Zueco}}, \bibinfo {author} {\bibfnamefont {J.~J.}\ \bibnamefont {Mazo}},
  \bibinfo {author} {\bibfnamefont {E.}~\bibnamefont {Solano}}, \ and\ \bibinfo
  {author} {\bibfnamefont {J.~J.}\ \bibnamefont {Garc\'{\i}a-Ripoll}},\ }\href
  {\doibase 10.1103/PhysRevB.86.024503} {\bibfield  {journal} {\bibinfo
  {journal} {Phys. Rev. B}\ }\textbf {\bibinfo {volume} {86}},\ \bibinfo
  {pages} {024503} (\bibinfo {year} {2012})}\BibitemShut {NoStop}%
\bibitem [{\citenamefont {Xu}\ \emph {et~al.}(2000)\citenamefont {Xu},
  \citenamefont {Li}, \citenamefont {Lee},\ and\ \citenamefont
  {Yariv}}]{Xu-PRE2000}%
  \BibitemOpen
  \bibfield  {author} {\bibinfo {author} {\bibfnamefont {Y.}~\bibnamefont
  {Xu}}, \bibinfo {author} {\bibfnamefont {Y.}~\bibnamefont {Li}}, \bibinfo
  {author} {\bibfnamefont {R.~K.}\ \bibnamefont {Lee}}, \ and\ \bibinfo
  {author} {\bibfnamefont {A.}~\bibnamefont {Yariv}},\ }\href {\doibase
  10.1103/PhysRevE.62.7389} {\bibfield  {journal} {\bibinfo  {journal} {Phys.
  Rev. E}\ }\textbf {\bibinfo {volume} {62}},\ \bibinfo {pages} {7389}
  (\bibinfo {year} {2000})}\BibitemShut {NoStop}%
\bibitem [{\citenamefont {Shen}\ and\ \citenamefont
  {Fan}(2005{\natexlab{b}})}]{Shen-OL2005}%
  \BibitemOpen
  \bibfield  {author} {\bibinfo {author} {\bibfnamefont {J.-T.}\ \bibnamefont
  {Shen}}\ and\ \bibinfo {author} {\bibfnamefont {S.}~\bibnamefont {Fan}},\
  }\href {\doibase 10.1364/OL.30.002001} {\bibfield  {journal} {\bibinfo
  {journal} {Opt. Lett.}\ }\textbf {\bibinfo {volume} {30}},\ \bibinfo {pages}
  {2001} (\bibinfo {year} {2005}{\natexlab{b}})}\BibitemShut {NoStop}%
\bibitem [{\citenamefont {Yanik}\ \emph {et~al.}(2004)\citenamefont {Yanik},
  \citenamefont {Suh}, \citenamefont {Wang},\ and\ \citenamefont
  {Fan}}]{Yanik-PRL2004}%
  \BibitemOpen
  \bibfield  {author} {\bibinfo {author} {\bibfnamefont {M.~F.}\ \bibnamefont
  {Yanik}}, \bibinfo {author} {\bibfnamefont {W.}~\bibnamefont {Suh}}, \bibinfo
  {author} {\bibfnamefont {Z.}~\bibnamefont {Wang}}, \ and\ \bibinfo {author}
  {\bibfnamefont {S.}~\bibnamefont {Fan}},\ }\href {\doibase
  10.1103/PhysRevLett.93.233903} {\bibfield  {journal} {\bibinfo  {journal}
  {Phys. Rev. Lett.}\ }\textbf {\bibinfo {volume} {93}},\ \bibinfo {pages}
  {233903} (\bibinfo {year} {2004})}\BibitemShut {NoStop}%
\bibitem [{\citenamefont {Shen}\ \emph {et~al.}(2007)\citenamefont {Shen},
  \citenamefont {Povinelli}, \citenamefont {Sandhu},\ and\ \citenamefont
  {Fan}}]{Shen-PRB2007}%
  \BibitemOpen
  \bibfield  {author} {\bibinfo {author} {\bibfnamefont {J.-T.}\ \bibnamefont
  {Shen}}, \bibinfo {author} {\bibfnamefont {M.~L.}\ \bibnamefont {Povinelli}},
  \bibinfo {author} {\bibfnamefont {S.}~\bibnamefont {Sandhu}}, \ and\ \bibinfo
  {author} {\bibfnamefont {S.}~\bibnamefont {Fan}},\ }\href {\doibase
  10.1103/PhysRevB.75.035320} {\bibfield  {journal} {\bibinfo  {journal} {Phys.
  Rev. B}\ }\textbf {\bibinfo {volume} {75}},\ \bibinfo {pages} {035320}
  (\bibinfo {year} {2007})}\BibitemShut {NoStop}%
\bibitem [{\citenamefont {Yi}\ \emph {et~al.}(2010)\citenamefont {Yi},
  \citenamefont {Citrin},\ and\ \citenamefont {Zhou}}]{Yi-OE2010}%
  \BibitemOpen
  \bibfield  {author} {\bibinfo {author} {\bibfnamefont {H.}~\bibnamefont
  {Yi}}, \bibinfo {author} {\bibfnamefont {D.~S.}\ \bibnamefont {Citrin}}, \
  and\ \bibinfo {author} {\bibfnamefont {Z.}~\bibnamefont {Zhou}},\ }\href
  {\doibase 10.1364/OE.18.002967} {\bibfield  {journal} {\bibinfo  {journal}
  {Opt. Express}\ }\textbf {\bibinfo {volume} {18}},\ \bibinfo {pages} {2967}
  (\bibinfo {year} {2010})}\BibitemShut {NoStop}%
\bibitem [{\citenamefont {Witthaut}\ and\ \citenamefont
  {Sørensen}(2010)}]{Witthaut-NJP2010}%
  \BibitemOpen
  \bibfield  {author} {\bibinfo {author} {\bibfnamefont {D.}~\bibnamefont
  {Witthaut}}\ and\ \bibinfo {author} {\bibfnamefont {A.~S.}\ \bibnamefont
  {Sørensen}},\ }\href {\doibase 10.1088/1367-2630/12/4/043052} {\bibfield
  {journal} {\bibinfo  {journal} {New J. Phys.}\ }\textbf {\bibinfo {volume}
  {12}},\ \bibinfo {pages} {043052} (\bibinfo {year} {2010})}\BibitemShut
  {NoStop}%
\bibitem [{\citenamefont {Fang}\ and\ \citenamefont
  {Baranger}(2015)}]{Fang-PRA2015}%
  \BibitemOpen
  \bibfield  {author} {\bibinfo {author} {\bibfnamefont {Y.-L.~L.}\
  \bibnamefont {Fang}}\ and\ \bibinfo {author} {\bibfnamefont {H.~U.}\
  \bibnamefont {Baranger}},\ }\href {\doibase 10.1103/PhysRevA.91.053845}
  {\bibfield  {journal} {\bibinfo  {journal} {Phys. Rev. A}\ }\textbf {\bibinfo
  {volume} {91}},\ \bibinfo {pages} {053845} (\bibinfo {year}
  {2015})}\BibitemShut {NoStop}%
\bibitem [{\citenamefont {Mirza}\ \emph {et~al.}(2017)\citenamefont {Mirza},
  \citenamefont {Hoskins},\ and\ \citenamefont {Schotland}}]{Mirza-PRA2017}%
  \BibitemOpen
  \bibfield  {author} {\bibinfo {author} {\bibfnamefont {I.~M.}\ \bibnamefont
  {Mirza}}, \bibinfo {author} {\bibfnamefont {J.~G.}\ \bibnamefont {Hoskins}},
  \ and\ \bibinfo {author} {\bibfnamefont {J.~C.}\ \bibnamefont {Schotland}},\
  }\href {\doibase 10.1103/PhysRevA.96.053804} {\bibfield  {journal} {\bibinfo
  {journal} {Phys. Rev. A}\ }\textbf {\bibinfo {volume} {96}},\ \bibinfo
  {pages} {053804} (\bibinfo {year} {2017})}\BibitemShut {NoStop}%
\bibitem [{\citenamefont {Mirhosseini}\ \emph {et~al.}(2018)\citenamefont
  {Mirhosseini}, \citenamefont {Kim}, \citenamefont {Ferreira}, \citenamefont
  {Kalaee}, \citenamefont {Sipahigil}, \citenamefont {Keller},\ and\
  \citenamefont {Painter}}]{Mohammad-Natcom2018}%
  \BibitemOpen
  \bibfield  {author} {\bibinfo {author} {\bibfnamefont {M.}~\bibnamefont
  {Mirhosseini}}, \bibinfo {author} {\bibfnamefont {E.}~\bibnamefont {Kim}},
  \bibinfo {author} {\bibfnamefont {V.~S.}\ \bibnamefont {Ferreira}}, \bibinfo
  {author} {\bibfnamefont {M.}~\bibnamefont {Kalaee}}, \bibinfo {author}
  {\bibfnamefont {A.}~\bibnamefont {Sipahigil}}, \bibinfo {author}
  {\bibfnamefont {A.~J.}\ \bibnamefont {Keller}}, \ and\ \bibinfo {author}
  {\bibfnamefont {O.}~\bibnamefont {Painter}},\ }\href {\doibase
  10.1038/s41467-018-06142-z} {\bibfield  {journal} {\bibinfo  {journal} {Nat.
  Commun.}\ }\textbf {\bibinfo {volume} {9}},\ \bibinfo {pages} {3706}
  (\bibinfo {year} {2018})}\BibitemShut {NoStop}%
\bibitem [{\citenamefont {He}\ \emph {et~al.}(2021)\citenamefont {He},
  \citenamefont {He},\ and\ \citenamefont {Wei}}]{He-OE2021}%
  \BibitemOpen
  \bibfield  {author} {\bibinfo {author} {\bibfnamefont {S.}~\bibnamefont
  {He}}, \bibinfo {author} {\bibfnamefont {Q.}~\bibnamefont {He}}, \ and\
  \bibinfo {author} {\bibfnamefont {L.~F.}\ \bibnamefont {Wei}},\ }\href
  {\doibase 10.1364/OE.445444} {\bibfield  {journal} {\bibinfo  {journal} {Opt.
  Express}\ }\textbf {\bibinfo {volume} {29}},\ \bibinfo {pages} {43148}
  (\bibinfo {year} {2021})}\BibitemShut {NoStop}%
\bibitem [{\citenamefont {Greenberg}\ \emph {et~al.}(2021)\citenamefont
  {Greenberg}, \citenamefont {Shtygashev},\ and\ \citenamefont
  {Moiseev}}]{Greenberg-PRA2021}%
  \BibitemOpen
  \bibfield  {author} {\bibinfo {author} {\bibfnamefont {Y.~S.}\ \bibnamefont
  {Greenberg}}, \bibinfo {author} {\bibfnamefont {A.~A.}\ \bibnamefont
  {Shtygashev}}, \ and\ \bibinfo {author} {\bibfnamefont {A.~G.}\ \bibnamefont
  {Moiseev}},\ }\href {\doibase 10.1103/PhysRevA.103.023508} {\bibfield
  {journal} {\bibinfo  {journal} {Phys. Rev. A}\ }\textbf {\bibinfo {volume}
  {103}},\ \bibinfo {pages} {023508} (\bibinfo {year} {2021})}\BibitemShut
  {NoStop}%
\bibitem [{\citenamefont {Tang}\ \emph {et~al.}(2022)\citenamefont {Tang},
  \citenamefont {Nie}, \citenamefont {Tang}, \citenamefont {Chen},
  \citenamefont {Su}, \citenamefont {Lu}, \citenamefont {Nori},\ and\
  \citenamefont {Xia}}]{Tang-PRL2022}%
  \BibitemOpen
  \bibfield  {author} {\bibinfo {author} {\bibfnamefont {J.-S.}\ \bibnamefont
  {Tang}}, \bibinfo {author} {\bibfnamefont {W.}~\bibnamefont {Nie}}, \bibinfo
  {author} {\bibfnamefont {L.}~\bibnamefont {Tang}}, \bibinfo {author}
  {\bibfnamefont {M.}~\bibnamefont {Chen}}, \bibinfo {author} {\bibfnamefont
  {X.}~\bibnamefont {Su}}, \bibinfo {author} {\bibfnamefont {Y.}~\bibnamefont
  {Lu}}, \bibinfo {author} {\bibfnamefont {F.}~\bibnamefont {Nori}}, \ and\
  \bibinfo {author} {\bibfnamefont {K.}~\bibnamefont {Xia}},\ }\href {\doibase
  10.1103/PhysRevLett.128.203602} {\bibfield  {journal} {\bibinfo  {journal}
  {Phys. Rev. Lett.}\ }\textbf {\bibinfo {volume} {128}},\ \bibinfo {pages}
  {203602} (\bibinfo {year} {2022})}\BibitemShut {NoStop}%
\bibitem [{\citenamefont {Peng}\ and\ \citenamefont
  {Jia}(2023)}]{Peng-PRA2023}%
  \BibitemOpen
  \bibfield  {author} {\bibinfo {author} {\bibfnamefont {Y.~P.}\ \bibnamefont
  {Peng}}\ and\ \bibinfo {author} {\bibfnamefont {W.~Z.}\ \bibnamefont {Jia}},\
  }\href {\doibase 10.1103/PhysRevA.108.043709} {\bibfield  {journal} {\bibinfo
   {journal} {Phys. Rev. A}\ }\textbf {\bibinfo {volume} {108}},\ \bibinfo
  {pages} {043709} (\bibinfo {year} {2023})}\BibitemShut {NoStop}%
\bibitem [{\citenamefont {Berndsen}\ and\ \citenamefont
  {Mirza}(2023)}]{Berndsen-PRA2023}%
  \BibitemOpen
  \bibfield  {author} {\bibinfo {author} {\bibfnamefont {T.}~\bibnamefont
  {Berndsen}}\ and\ \bibinfo {author} {\bibfnamefont {I.~M.}\ \bibnamefont
  {Mirza}},\ }\href {\doibase 10.1103/PhysRevA.108.063702} {\bibfield
  {journal} {\bibinfo  {journal} {Phys. Rev. A}\ }\textbf {\bibinfo {volume}
  {108}},\ \bibinfo {pages} {063702} (\bibinfo {year} {2023})}\BibitemShut
  {NoStop}%
\bibitem [{\citenamefont {Berndsen}\ \emph {et~al.}(2024)\citenamefont
  {Berndsen}, \citenamefont {Amgain},\ and\ \citenamefont
  {Mirza}}]{Berndsen-JOSAB2024}%
  \BibitemOpen
  \bibfield  {author} {\bibinfo {author} {\bibfnamefont {T.}~\bibnamefont
  {Berndsen}}, \bibinfo {author} {\bibfnamefont {N.}~\bibnamefont {Amgain}}, \
  and\ \bibinfo {author} {\bibfnamefont {I.}~\bibnamefont {Mirza}},\ }\href
  {\doibase 10.1364/JOSAB.520000} {\bibfield  {journal} {\bibinfo  {journal}
  {J. Opt. Soc. Am. B}\ }\textbf {\bibinfo {volume} {41}},\ \bibinfo {pages}
  {C9} (\bibinfo {year} {2024})}\BibitemShut {NoStop}%
\bibitem [{\citenamefont {Hime}\ \emph {et~al.}(2006)\citenamefont {Hime},
  \citenamefont {Reichardt}, \citenamefont {Plourde}, \citenamefont
  {Robertson}, \citenamefont {Wu}, \citenamefont {Ustinov},\ and\ \citenamefont
  {Clarke}}]{Hime-Science2006}%
  \BibitemOpen
  \bibfield  {author} {\bibinfo {author} {\bibfnamefont {T.}~\bibnamefont
  {Hime}}, \bibinfo {author} {\bibfnamefont {P.~A.}\ \bibnamefont {Reichardt}},
  \bibinfo {author} {\bibfnamefont {B.~L.~T.}\ \bibnamefont {Plourde}},
  \bibinfo {author} {\bibfnamefont {T.~L.}\ \bibnamefont {Robertson}}, \bibinfo
  {author} {\bibfnamefont {C.-E.}\ \bibnamefont {Wu}}, \bibinfo {author}
  {\bibfnamefont {A.~V.}\ \bibnamefont {Ustinov}}, \ and\ \bibinfo {author}
  {\bibfnamefont {J.}~\bibnamefont {Clarke}},\ }\href {\doibase
  10.1126/science.1134388} {\bibfield  {journal} {\bibinfo  {journal}
  {Science}\ }\textbf {\bibinfo {volume} {314}},\ \bibinfo {pages} {1427}
  (\bibinfo {year} {2006})}\BibitemShut {NoStop}%
\bibitem [{\citenamefont {Niskanen}\ \emph {et~al.}(2007)\citenamefont
  {Niskanen}, \citenamefont {Harrabi}, \citenamefont {Yoshihara}, \citenamefont
  {Nakamura}, \citenamefont {Lloyd},\ and\ \citenamefont
  {Tsai}}]{Niskanen-Science2007}%
  \BibitemOpen
  \bibfield  {author} {\bibinfo {author} {\bibfnamefont {A.~O.}\ \bibnamefont
  {Niskanen}}, \bibinfo {author} {\bibfnamefont {K.}~\bibnamefont {Harrabi}},
  \bibinfo {author} {\bibfnamefont {F.}~\bibnamefont {Yoshihara}}, \bibinfo
  {author} {\bibfnamefont {Y.}~\bibnamefont {Nakamura}}, \bibinfo {author}
  {\bibfnamefont {S.}~\bibnamefont {Lloyd}}, \ and\ \bibinfo {author}
  {\bibfnamefont {J.~S.}\ \bibnamefont {Tsai}},\ }\href {\doibase
  10.1126/science.1141324} {\bibfield  {journal} {\bibinfo  {journal}
  {Science}\ }\textbf {\bibinfo {volume} {316}},\ \bibinfo {pages} {723}
  (\bibinfo {year} {2007})}\BibitemShut {NoStop}%
\bibitem [{\citenamefont {Baust}\ \emph {et~al.}(2015)\citenamefont {Baust},
  \citenamefont {Hoffmann}, \citenamefont {Haeberlein}, \citenamefont
  {Schwarz}, \citenamefont {Eder}, \citenamefont {Goetz}, \citenamefont
  {Wulschner}, \citenamefont {Xie}, \citenamefont {Zhong}, \citenamefont
  {Quijandr\'{\i}a}, \citenamefont {Peropadre}, \citenamefont {Zueco},
  \citenamefont {Garc\'{\i}a~Ripoll}, \citenamefont {Solano}, \citenamefont
  {Fedorov}, \citenamefont {Menzel}, \citenamefont {Deppe}, \citenamefont
  {Marx},\ and\ \citenamefont {Gross}}]{Baust-PRB2015}%
  \BibitemOpen
  \bibfield  {author} {\bibinfo {author} {\bibfnamefont {A.}~\bibnamefont
  {Baust}}, \bibinfo {author} {\bibfnamefont {E.}~\bibnamefont {Hoffmann}},
  \bibinfo {author} {\bibfnamefont {M.}~\bibnamefont {Haeberlein}}, \bibinfo
  {author} {\bibfnamefont {M.~J.}\ \bibnamefont {Schwarz}}, \bibinfo {author}
  {\bibfnamefont {P.}~\bibnamefont {Eder}}, \bibinfo {author} {\bibfnamefont
  {J.}~\bibnamefont {Goetz}}, \bibinfo {author} {\bibfnamefont
  {F.}~\bibnamefont {Wulschner}}, \bibinfo {author} {\bibfnamefont
  {E.}~\bibnamefont {Xie}}, \bibinfo {author} {\bibfnamefont {L.}~\bibnamefont
  {Zhong}}, \bibinfo {author} {\bibfnamefont {F.}~\bibnamefont
  {Quijandr\'{\i}a}}, \bibinfo {author} {\bibfnamefont {B.}~\bibnamefont
  {Peropadre}}, \bibinfo {author} {\bibfnamefont {D.}~\bibnamefont {Zueco}},
  \bibinfo {author} {\bibfnamefont {J.-J.}\ \bibnamefont {Garc\'{\i}a~Ripoll}},
  \bibinfo {author} {\bibfnamefont {E.}~\bibnamefont {Solano}}, \bibinfo
  {author} {\bibfnamefont {K.}~\bibnamefont {Fedorov}}, \bibinfo {author}
  {\bibfnamefont {E.~P.}\ \bibnamefont {Menzel}}, \bibinfo {author}
  {\bibfnamefont {F.}~\bibnamefont {Deppe}}, \bibinfo {author} {\bibfnamefont
  {A.}~\bibnamefont {Marx}}, \ and\ \bibinfo {author} {\bibfnamefont
  {R.}~\bibnamefont {Gross}},\ }\href {\doibase 10.1103/PhysRevB.91.014515}
  {\bibfield  {journal} {\bibinfo  {journal} {Phys. Rev. B}\ }\textbf {\bibinfo
  {volume} {91}},\ \bibinfo {pages} {014515} (\bibinfo {year}
  {2015})}\BibitemShut {NoStop}%
\bibitem [{\citenamefont {Zhu}\ and\ \citenamefont {Jia}(2019)}]{Zhu-PRA2019}%
  \BibitemOpen
  \bibfield  {author} {\bibinfo {author} {\bibfnamefont {Y.~T.}\ \bibnamefont
  {Zhu}}\ and\ \bibinfo {author} {\bibfnamefont {W.~Z.}\ \bibnamefont {Jia}},\
  }\href {\doibase 10.1103/PhysRevA.99.063815} {\bibfield  {journal} {\bibinfo
  {journal} {Phys. Rev. A}\ }\textbf {\bibinfo {volume} {99}},\ \bibinfo
  {pages} {063815} (\bibinfo {year} {2019})}\BibitemShut {NoStop}%
\bibitem [{\citenamefont {{Abelès, Florin}}(1950)}]{Abeles-AnnPhys1950}%
  \BibitemOpen
  \bibfield  {author} {\bibinfo {author} {\bibnamefont {{Abelès, Florin}}},\
  }\href {\doibase 10.1051/anphys/195012050596} {\bibfield  {journal} {\bibinfo
   {journal} {Ann. Phys.}\ }\textbf {\bibinfo {volume} {12}},\ \bibinfo {pages}
  {596} (\bibinfo {year} {1950})}\BibitemShut {NoStop}%
\bibitem [{\citenamefont {Brody}(2013)}]{Brody-JPA2013}%
  \BibitemOpen
  \bibfield  {author} {\bibinfo {author} {\bibfnamefont {D.~C.}\ \bibnamefont
  {Brody}},\ }\href {\doibase 10.1088/1751-8113/47/3/035305} {\bibfield
  {journal} {\bibinfo  {journal} {J. Phys. A-Math. Theor.}\ }\textbf {\bibinfo
  {volume} {47}},\ \bibinfo {pages} {035305} (\bibinfo {year}
  {2013})}\BibitemShut {NoStop}%
\end{thebibliography}%
\end{document}